\documentclass[a4paper,6pt,intlimits]{iopart}
\pdfoutput=1

\expandafter\let\csname equation*\endcsname\relax
\expandafter\let\csname endequation*\endcsname\relax

\usepackage{amsmath}
\usepackage{amsthm,amssymb,mathrsfs,xfrac}
\usepackage{cite}
\usepackage[caption]{subfig}
\usepackage[english]{babel}
\usepackage{pgfplotstable}
\usepackage{hyperref}

\def\be{\begin{equation}}
\def\ee{\end{equation}}

\def\bea#1\eea{\begin{align}#1\end{align}}

\newcommand{\mO}{\mathcal{O}}

\newcommand{\mP}{\mathcal{P}}

\newcommand{\mC}{\mathcal{C}}
\newcommand{\mI}{\mathcal{I}}
\newcommand{\mL}{\mathcal{L}}

\newcommand{\ts}{\tilde{s}}

\usepackage{color}
\newcommand{\red}{\color{black}}
\newcommand{\blue}{\color{black}}

\newcommand{\mycol}[1]{{\color{black} #1}}

\usepackage{bm}% bo

\begin{document}

\title[A First-Order Dynamical Transition in the displacement
  distribution of a Driven Run-and-Tumble Particle]{A First-Order
  Dynamical Transition in the displacement distribution of a Driven
  Run-and-Tumble Particle}

\date{\today}

%\iffalse
\author{Giacomo Gradenigo}%%\ead{ggradenigo@gmail.com}
\address{Dipartimento di Fisica, Universit\`a Sapienza, Piazzale Aldo
  Moro 5, I-00185, Rome, Italy} \address{CNR-Nanotec, Institute of
  Nanotechnology, UOS-Roma, Rome, Italy}

\author{Satya N. Majumdar}%%\ead{majumdar@lptms.u-psud.fr}
\address{LPTMS, CNRS, Univ. Paris-Sud, Universit\'e Paris-Saclay,
  91405 Orsay, France}

%%%%%%%%%%%%%%%%%%%%%%%%%%%%%%%%%%%%%%%%%%%%%%%%%%%%%%%%%%%%%%%%%%%%%%%%
%%%%%%%%%%%%%%%%%%%%%%%%%%%%%%%%%%%%%%%%%%%%%%%%%%%%%%%%%%%%%%%%%%%%%%%%
%%
\begin{abstract}
We study the probability distribution $P(X_N=X,N)$ of the total
displacement $X_N$ of an $N$-step run and tumble particle on a line,
in presence of a constant nonzero drive $E$. While the central limit
theorem predicts a standard Gaussian form for $P(X,N)$ near its peak,
we show that for large positive and negative $X$, the distribution
exhibits anomalous large deviation forms. For large positive $X$, the
associated rate function is nonanalytic at a critical value of the
scaled distance from the peak where its first derivative is
discontinuous. This signals a first-order dynamical phase transition
from a homogeneous `fluid' phase to a `condensed' phase that is
dominated by a single large run. A similar first-order transition
occurs for negative large fluctuations as well. Numerical simulations
are in excellent agreement with our analytical predictions.
\end{abstract}

\maketitle
%\tableofcontents

%%%%%%%%%%%%%%%%%%%%%%%%%%%%%%%%%%%%%%%%%%%
%%%%%%%%%%%%%%%%% INTRO %%%%%%%%%%%%%%%%%%%%%
%%%%%%%%%%%%%%%%%%%%%%

\section{Introduction}
\label{sec:intro}

Recent years have seen immense theoretical and experimental interest
in the study of `active' systems, consisting of self-propelled
individual particles~\cite{Marchetti1,Bechinger,Ramaswamy,Fodor}. 
These active particles exhibit novel collective
nonequilibrium pheomena, such as the motility induced phase separation
(MIPS)~\cite{TC08,FM12,BBKLBS13,CT13,CT15,SCT15,B03}, clustering
effect~\cite{Slowman}, spontaneous segregation of mixtures of active
and passive particles~\cite{Stenhammer} and many other interesting
effects. {\red{ These collective effects arise from a combination of
self-propulsion and interaction between the active particles. However,
even in the absence of interactions between particles (noninteracting limit),
the stochastic process associated with a single active particle is
rather interesting due purely to the self-propulsion. This self-propulsion
induces a memory or `persistence' in the effective noise felt by the particle, leading often
to interesting non-Markovian effects. At
the level of individual particles, the simplest examples of such
active particles are the so called active Brownian motion (ABM) or the
`Run-and-Tumble' particle (RTP) [for a recent pedagogical review,
see~\cite{Fodor}]. For a single ABM particle, both free as well as confined in a harmonic trap, 
there have been a number of recent theoretical
and experimental studies in two dimensions on the 
position distribution~\cite{SG14,SS15,Takatori16,Kurz18,BMRS18,DD18,ICTS19}, 
as well as on its first-passage properties~\cite{BMRS18}. In this paper, we will focus on the
other well studied model of a single active particle, namely the RTP~\cite{TC08,CT13,SCT15} in
one dimension, but subjected to a constant external force $E>0$. 

RTP process mimicks the typical
motion of bacterias such as E. Coli~\cite{B03}: a particle moves ballistically at a constant speed for
random durations of time, called ``runs'', until sudden changes
of direction and speed take place, called ``tumbles''. The tumbling occurs as a Poisson
process with rate $\gamma$, i.e., the distribution of the duration of a single run between two successive tumblings
is exponential with parameter $\gamma$. We will set $\gamma=1$ for the rest of the paper.
We will focus here in one dimension. At the end of each tumbling, the particle chooses a new velocity
drawn independently (from run to run) from a probability distribution funtion (PDF) $q(v)$, which
is typically symmetric. In the standard RTP model known as the persistent random walk, $q(v)$ is chosen to be bimodal:
\be
q(v)= \frac{1}{2}\, \left[\delta(v-v_0)+ \delta(v+v_0)\right]\, . 
\label{bimodal.1}
\ee
In this model, the position $x(t)$ of the RTP evolves in time
via
\begin{equation}
\frac{dx}{dt}= v_0\, \sigma(t)\, ,
\label{prw.1}
\end{equation}
where $\sigma(t)=\pm 1$ is a dichotomous telegraphic noise that flips
between the two states with a constant rate $\gamma=1$.  This
persistent random walk model has been studied extensively in the past
and many properties are known, e.g. the propagator and the mean exit
time from a finite interval, amongst other
observables~\cite{ML_2017,Weiss_2002}. In one dimension, there have
been a number of recent theoretical studies on the first-passage
properties of a free
RTP~\cite{ADP_2014,LA_2015,Malakar_2018,DM_2018,EM_2018,LMS_2019} and,
very recently, for an RTP subjected to an external confining
potential~\cite{Dhar_18}.

In this paper, we will study a variant of this standard RTP model in one dimension. In our model, while the duration of a 
run, say $\tau_i$ for the $i$-th run, is still exponentially
distributed with rate $\gamma=1$, the actual motion during a `run' is different. In our model there
is an external force $E>0$ that drives the particle during a run. More precisely, at the begining of the $i$-th run, the
particle again chooses a new velocity $v_i$ from a PDF $q(v)$ (which is not necessarily bimodal). Then starting
with this initial velocity $v_i$, the particle moves via Newton's second law during the run duration $0\le t\le \tau_i$
\begin{equation}
\frac{dx}{dt}= v(t);\, \quad\quad m\, \frac{dv}{dt}= E\, ;  \quad v(0)=v_i\, .
\label{model.0}
\end{equation}
Thus we assume that there is no friction due to the environment (the particle's motion is thus not overdamped
as in standard Brownian motion). We will also set the mass $m=1$ for simplicity. Integrating Eq. (\ref{model.0}) trivially,
the displacement $x_i$ during the $i$-th run is given by
\begin{equation}
x_i= v_i \tau_i + \frac{E}{2}\, \tau_i^2 \, 
\label{model.1}
\end{equation}
where both $\tau_i$ and $v_i$ are independent random variables, drawn
respectively from $p(\tau)=e^{-\tau}\, \theta(\tau)$ and $q(v)$ which
is arbitrary (albeit symmetric). For $E=0$ and $q(v)$ bimodal as in
Eq. (\ref{bimodal.1}), our model reduces to the standard RTP.  We will
focus here on $E>0$ and Gaussian velocity distribution $q(v)=
e^{-v^2/2}/\sqrt{2\pi}$, though our results on condensation (see
later) will hold for a large class of velocity distributions $q(v)$,
including the bimodel case discussed above. We consider $N$
successive runs. We work here in the ensemble where the total number
$N$ of runs (or tumbles) is fixed, rather than the total time elapsed
$t$. However, our results can be easily extended to constant time
ensemble.  In the presence of a constant force $E$, the total distance
travelled by the particle after $N$ runs is
\be X_N= \sum_{i=1}^N x_i= \sum_{i=1}^N \left[v_i \tau_i + \frac{E}{2} \tau_i^2\right]
\label{dist_E.1}
\ee
In this paper, we are interested in the PDF $P(X, N)= {\rm
  Prob.}[X_N=X]$ of the total displacement for large $N$. Our main new
result is that for $E>0$ and for a broad class of $q(v)$'s including
the Gaussian and the bimodal distributions, the PDF $P(X,N)$, for
large $N$, exhibits a non-analyticity as a function of
$X$---signalling a condenation type first-order `phase transition' in
the system, as disussed below. We show that for the standard RTP, i.e,
for $E=0$ and $q(v)$ bimodal as in Eq. (\ref{bimodal.1}), this
interesting phase transition disappears. Our main results are
summarized in the next section. Below we discuss some qualitative
features of this PDF $P(X,N)$ for large $N$ and the physics behind the
phase transition, before moving to a more quantitative detailed
analysis in the later sections.

}}

\begin{figure}
\includegraphics[width=\columnwidth]{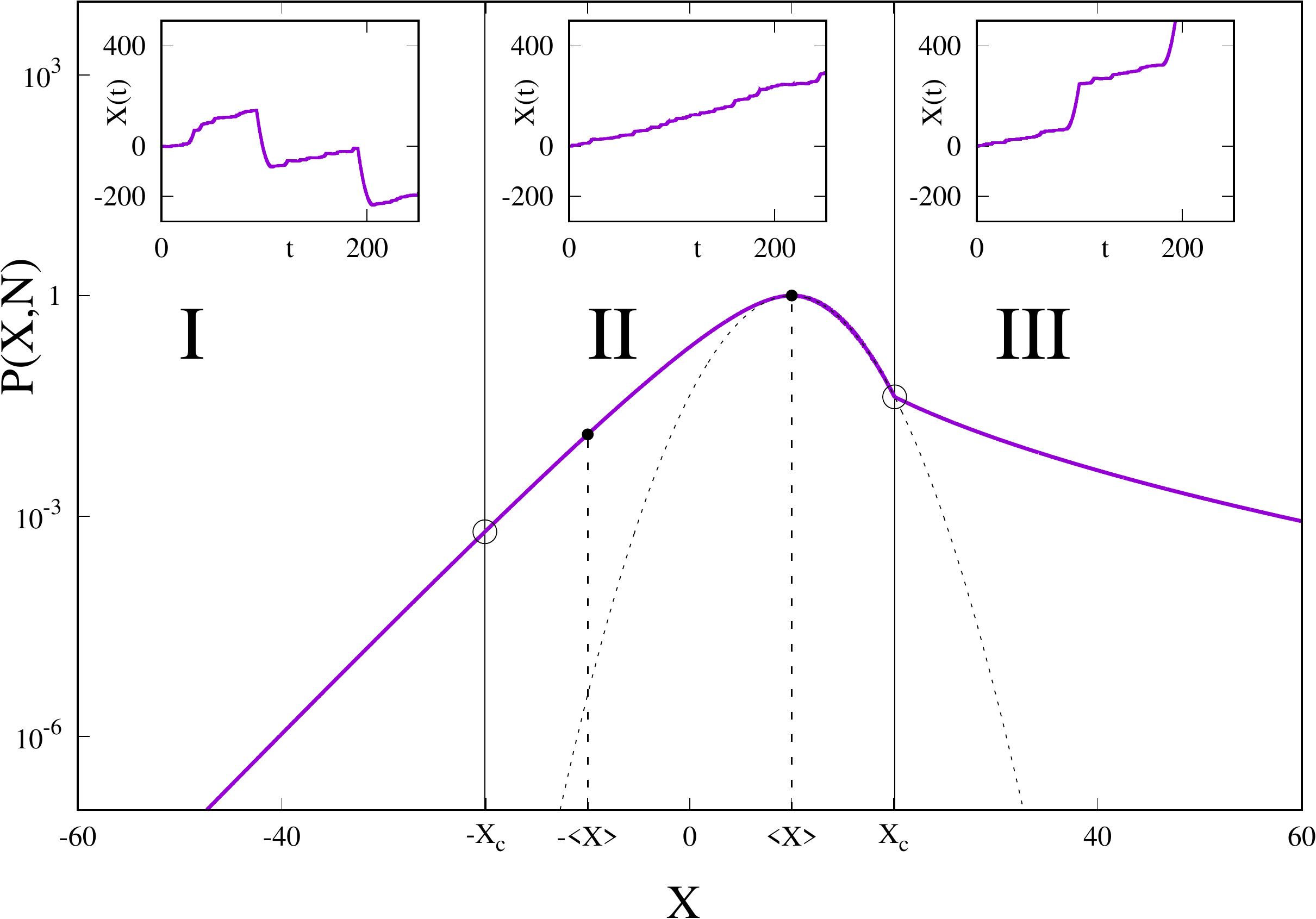}
\caption{\emph{Main}: Schematic representation of $P(X,N)$. Mean value
  is $\langle X\rangle = E\, N$.  Vertical lines separates three
  regions, $I$, $II$ and $III$, corresponding respectively to
  $X<-X_c$, $-X_c\le X< X_c$ and $X>X_c$. The two dynamical
  transitions are located at $X_c$ and $-X_c$. Region $II$ is the
  homogeneous phase. Regions $I$ and $III$ are the condensed
  phases. Dotted parabola is a guide to the eye. For $X\in [-\langle X
    \rangle,\langle X\rangle]$ (inside the dotted vertical lines), the
  PDF $P(X,N)$ is the result of a saddle-point approximation.
  \emph{Insets}: typical trajectories for each of the three regions:
  homogeneous trajectories for $II$, dominated by one single run for
  $I$ and $III$.}
\label{fig:cartoon}
\end{figure}

Due to the presence of a nonzero $E>0$, the PDF $P(X,N)$ is clearly asymmetric
as a fuction of $X$ (see Fig.~\ref{fig:cartoon}), and it has three
regimes which are denoted as $I$, $II$ and $III$ in
Fig.~\ref{fig:cartoon}.
In the central regime $II$, $P(X,N)$ describes the probability of {\em
  typical} fluctuations of $X$, while regimes $I$ and $III$ correspond
to {\em atypically} large fluctuations of $X$ on the negative and the
positive side respectively. A ``kink'', which is shown schematically
in Fig. \ref{fig:cartoon} at $X=X_c$, separates the typical
fluctuations regime ($II$) from positive large deviations (regime
$III$).  Another similar kink at $X=-X_c$ (see
Fig. (\ref{fig:cartoon})) separates regimes $I$ and $II$ on the side
of negative fluctuations.  The main result of this paper is to
demonstrate that this change in the nature of the fluctuations at
$X=X_c$ corresponds to a dynamical first-order transition. A similar
transition separates fluctuations in the center of the distribution
from {\em negative} large deviations (at the second kink on the left
at $X=-X_c$, although in this case the transition is ``hidden'' by an
exponential prefactor $e^{-E|X|}$. In the central fluid regime $II$,
the total distance $X$ is democratically distributed between $N$
individual runs of average sizes, while in regime $III$ (respectively
in $I$) there is a large positive (negative) ``condensate'' (i.e., a
single run that is large) that coexists with $(N-1)$ typical runs. By
zooming in close to the kinks or the critical points, we show that
$P(X,N)$ is described by an {\em anomalous} large deviation form near
the kinks, with a {\em local} rate function that is continuous at the
kink, but its first-derivative has a discontinuous jump (see the
results of numerical simulations in Figs. \ref{fig:numerics})---thus
signalling a first-order phase transition.

Thus in our simple model, the PDF of the total displacement $P(X,N)$ exhibits a first-order phase transition, similar
to the condensation transition that occurs in
various lattice models of
mass-transport~\cite{MEZ05,EMZ06,EH05,S08-leshouches,SEM14,SEM14b,SEM17,GB17}. In
these models, each site of a lattice (of $N$ sites) has a certain mass
$m_i\ge 0$ and a fraction of the mass at each site gets transported to
a neighbouring site with a rate that depends on the local mass---total
mass $M$ is conserved by the dynamics. The Zero Range Process (ZRP) is
a special case of these more general mass-transport
models~\cite{MEZ05,EMZ06,EH05,S08-leshouches}.  The dynamics drives
the system into a nonequilibrium steady state and there is a whole
class of models for which the steady state has a product measure,
i.e., the joint distribution of masses in the steady state becomes
factorised~\cite{EH05}. Furthermore, under certain conditions, the
system in the steady state undergoes a phase transition from a fluid
phase for $M<M_c$, to a condensed phase $(M>M_c)$ where one single
site acquires a mass proportional to the total mass
$M$~\cite{EH05,S08-leshouches}.  This single site is the so called
condensate.
 
The total distance $X$ travelled by an RTP in $N$ runs in our model is
the counterpart to the total mass $M$ in mass-transport models on a
lattice with $N$ sites. Hence, the condensate (a single site carrying
a mass proportional to $N$) in the mass transport model corresponds to
a {\it single extensive run} in the RTP model.  One difference is that
in mass-transport models, the mass $M$ is always positive, unlike in
our case where $X$ can be both positive and negative. Consequently, we
have both ``positive'' and ``negative'' condensates, while in the
standard mass-transport models, there is only a ``positive''
condensate. This explains why we have a pair of critical points (see
Figs.~\ref{fig:cartoon}), as opposed to a single critical point in
mass-transport models.  Another difference lies in the observable of
interest. In mass-transport models, the central object of interest is
the mass distribution at a single lattice site and it shows different
behaviors across the condensation transition. In our case we focus on
a simpler object, the distribution $P(X,N)$ (playing the role of the
partition function in mass-transport models with a factorised steady
state), and we show how the signature of the condensation transition
is already manifest in $P(X,N)$ itself.  One of our main results is to
show that near this critical point, $P(X,N)$ exhibits an anomalous
large deviation form with an associated rate function that shows a
discontinuity in its first derivative.

The rest of the paper is organized as follows. In Section 2, we
define our model precisely and summarize the main results. Section 3
contains the most extensive analytical computation of the distribution $P(X,N)$
of the total displacement for $X>\langle X\rangle=EN$ (positive
fluctuations). {\red{It also includes a discussion on the details of numerical simulations (Section 3.4)}}.
Section 4 contains analogus computation of $P(X,N)$
for $X<-\langle X\rangle=-EN$, i.e, for negative fluctuations. Section
V contains a summary and conclusions. Finally, some details of the
computations are presented in the Appendices.

\section{The model and the summary of the main results}
\label{sec:model-results}

%%\subsection{ The model}

We consider a single RTP on a line, starting initially at $X=0$. Each
trajectory is made of $N$ independent runs. The $i$-th run starts with
initial velocity $v_i$ and lasts a random time $\tau_i$. The particle
is also subjected to a constant force (field) $E> 0$.  The total
displacement of the particle after $N$ runs, using Newton's law for
each run, is therefore given by
\be X_N=
\sum_{i=1}^N x_i= \sum_{i=1}^N \left[v_i \tau_i + \frac{E}{2}
  \tau_i^2\right] \, 
\label{eq:Xacc}
\ee
{\red{where $x_i$ denotes
the displacement during the $i$-th run.}}
The velocity $v_i$'s and the duration $\tau_i$'s for each run are i.i.d
random variables drawn from the normalized PDF's
\bea
q(v) &= \frac{1}{\sqrt{2\pi}} \exp\left[ -v^2/2\right] 
\label{vel_pdf} \\
p(\tau) &= \Theta(\tau)~\exp(-\tau)
\label{duration_pdf}
\eea where $\Theta(t)$ is the Heaviside theta function: $\Theta(t)=1$
for $t\ge 0$ and $\Theta(t)=0$ for $t<0$. {\red{Even though we present detailed results
only for the Gaussian velocity distribution $q(v)$ in Eq. (\ref{vel_pdf}), our main conclusions concerning 
the first-order phase transition is valid for a broad class of $q(v)$'s, including the
bimodal distribution in Eq. (\ref{bimodal.1}). }}
Our goal is to compute the
probability distribution $P(X,N)= {\rm Prob.}[X_N=X]$ of the total
displacement $X_N=\sum_{i=1}^N x_i$. Thus, $X_N$ is clearly 
a sum of $N$ i.i.d.
random variables. Each of 
the $x_i$'s
has the normalized marginal PDF
\be
\mP(x)= \int_{-\infty}^{\infty} dv~\int_0^{\infty} d\tau~q(v)\,p(\tau)\,
\delta(x- v \tau- E\tau^2/2)\, \,. 
\label{x_marg.1}
\ee
where $q(v)$ and $p(\tau)$ are given in Eqs. (\ref{vel_pdf}) and
(\ref{duration_pdf}) respectively. The mean and the variance
of the displacement during each run can be computed easily and one gets
\begin{eqnarray}
\langle x \rangle &= E \label{meanx} \\
\sigma^2 &=\langle x^2\rangle - {\langle x\rangle}^2= 2+ 5\, E^2
\label{varx}
\end{eqnarray} 
Computing explicitly $\mP(x)$ from Eq. (\ref{x_marg.1}) is hard, however
as we will see, what really matters for the large $N$ behavior of $P(X,N)$ is the
asymptotic tail behavior of $\mP(x)$. These tails can be explicitly
obtained (see Appendix A).
For large positive $x$ we get
\be
\mP(x\to \infty) \approx \frac{1}{E} e^{2/E^2}\, x^{-1/2}\, 
e^{-\sqrt{2x/E}}\ ,
\label{EP.2-text}
\ee
and for large negative $x$
%%%
\bea
\mP(x\to -\infty) &= e^{-E\, |x|} \mP(|x|) \nonumber \\
&\approx e^{-E\, |x|}\frac{e^{2/E^2}}{E} \, |x|^{-1/2}\, e^{-\sqrt{2|x|/E}}\ .
\label{EP.3-text}
\eea
%%
%%
%%%
Thus the PDF $P(X_N=X,N)$ of the sum $X_N=\sum_{i=1}^N x_i$ reads
\be
P(X,N) = \int_{-\infty}^{\infty} \left[\prod_{i=1}^N 
dx_i~\mP(x_i)\right]~\delta\left(X-\sum_{i=1}^N x_i\right).
\label{eq:PXN-free}
\ee
where $\mP(x)$ is given in Eq. (\ref{x_marg.1}).

\vskip 0.4cm

\noindent {\red{{\em Relation to mass transport models and a criterion for condensation transition.}}}
It is interesting to notice that $P(X,N)$ is formally similar to the
partition function of lattice models of mass-transport with a
factorised steady
state~\cite{MEZ05,EMZ06,EH05,S08-leshouches,SEM14,SEM14b}. The latter
reads as:
\be Z(M,N)=
\int_0^{\infty} \left[\prod_{i=1}^N dm_i~
  f(m_i)\right]~\delta\left(M-\sum_{i=1}^N m_i\right).
\label{eq:PF-ZRP}
\ee
where $m_i\ge 0$ denotes the mass at site $i$, $f(m_i)$ the
corresponding steady state weight and $M$ being the total
mass. Comparing Eqs.~(\ref{eq:PXN-free}) and~(\ref{eq:PF-ZRP}), and
identifying the run distance $x_i$ with the mass $m_i$, $X$ with $M$,
and $f(m_i)$ with $\mP(x_i)$, we see that formally, our $P(X,N)$ is
exactly the counterpart of the partition function $Z(M,N)$ in
mass-transport models: the only difference is that, at variance with
$m_i$'s which are non-negative variables, our $x_i$'s can be both
positive and negative, which give rise to two condensed phases 
(respectively with a long positive and a long negative run).\\
%%
%%
%%  TO BE DISPLACED AFTERWORDS
%%

{\red{Before discussing our strategy for the computation of $P(X,N)$ in Eq. (\ref{eq:PXN-free}) with
$\mP(x)$ given by (\ref{x_marg.1}), it is
useful to recall, from the literature on the mass-transport models, which classes of $\mP(x)$ may lead to
the phenomenon of condensation. For the mass-transport models with positive mass $m$ distributed via the
PDF $f(m)$ in Eq. (\ref{eq:PF-ZRP}), it is known~\cite{S08-leshouches} that a condensation occurs
when the tail of $f(m)$ remains bounded in the interval $e^{-c m}< f(m)< 1/m^2$, as $m\to \infty$, where $c>0$ is any 
positive constant.
The ZRP typically corresponds to $f(m)\sim m^{-1-\mu}$ with $\mu >1$, and hence exhibits condensation~\cite{EH05,S08-leshouches}.
However, another class of $f(m)$'s that satisfy these bounds for large $m$ is the so called
stretched exponential class:
$f(m) \sim \exp\left[-a\, m^{\alpha}\right]$ for large $m$, with $\alpha>0$. Hence this class will also
exhibit the condensation transition~\cite{SEM14,SEM14b}. In our RTP model with $p(\tau)=e^{-\tau}$
and $q(v)=e^{-v^2/2}/\sqrt{2\pi}$, we see that for large $x$, $\mP(x)$ in Eq. (\ref{EP.2-text}) decays
as a stretched exponential with the stretching exponent $\alpha=1/2$. Hence, we would expect a condensation
transition for large positive $X$. A similar argument on the negative side shows 
that we will have a condensation transition for large negative $X$ as well.
Hence, we expect that for any choice of $p(\tau)$ and $q(v)$ that leads via Eq. (\ref{x_marg.1}) to a
marginal distribution $\mP(x)$ which satisfies the bounds $e^{-c\, x}< \mP(x)< 1/x^2$ for large $x$, one will
get a condensation. For example, for $E>0$, $p(\tau)=e^{-\tau}\theta(\tau)$ and with a bimodal velocity 
distribution $q(v)$ as in Eq. (\ref{bimodal.1}),
it is easy to show (see Appendix A) that as $x\to \infty$, 
\be
\mP(x) \sim \frac{1}{\sqrt{2\,E\,x}}\, e^{-\sqrt{2x/E}}\, ,
\label{Ebimodal.1}
\ee
which again satisfies the criterion for condensation. However, for the standard RTP model, i.e., 
if $E=0$, $p(\tau)=e^{-\tau}$ and $q(v)$ is bimodal as in Eq. (\ref{bimodal.1}), one finds (see Appendix A)
\be
\mP(x)= \frac{1}{2v_0}\, e^{-|x|/v_0}\, , 
\label{E0bimodal.1}
\ee   
which does not satisfy the condensation criterion above. 
Hence, for the standard RTP, this condensation transition is absent.
Thus, we see that while we present detailed calaculations only for the Gaussian velocity distribution,
the phenomenon of condensation that we have found for the RTP model is robust: it occurs for a broad 
class of $p(\tau)$ and $q(v)$ that lead to
a marginal $\mP(x)$ satisfying the asymptotic bounds mentioned above. 
Incidentally, to the best of our knowledge, our model provides the first physical realization of the 
condensation belonging to this stretched exponential class. 

}}    

\vskip 0.4cm

\noindent {\red{{\em Strategy for the large $N$ analysis of $P(X,N)$.} 
Let us now briefly outline our strategy to compute analytically the PDF $P(X,N)$.}}
By using the integral representation of the delta function:
$\delta(X)= \int e^{s\, X}\, ds/(2\pi i)$, one can write $P(X,N)$ as
\bea P(X,N) &= \frac{1}{2\pi i} \int_{s_0-i\infty}^{s_0+i\infty}
ds~e^{N h(s)} \nonumber \\
h(s) &= sx+\log[\mL(s)]
\label{eq:inv-Lapl-transf}
\eea
where $x=X/N$ and \bea \mL(s) &= \int_{-\infty}^{\infty} dx\, e^{-s\,
  x}\, \mP(x)\, ,\nonumber \\ &= \sqrt{\pi}
\frac{e^{\frac{1}{2s(E-s)}}}{\sqrt{2s(E-s)}}
\textrm{erfc}\left[\frac{1}{\sqrt{2s(E-s)}}\right].
\label{eq:mL}
\eea
%% %%
where $\textrm{erfc}(z)= \frac{2}{\sqrt{\pi}}\, \int_{z}^\infty e^{-u^2}\, 
du$ is the complementary error function.

The integration contour in Eq.~(\ref{eq:inv-Lapl-transf}) is the
Bromwich contour in the complex $s$ plane. There are two possible
situations: (A) the equation $\partial h(s)/\partial s =0$ has a 
solution for real $s=s_0$ (see Fig. 3) and then the
the integral in
Eq.~(\ref{eq:inv-Lapl-transf}) can be computed for large $N$ using a 
saddle-point approximation; (B)
there is no saddle point and one has to carry out the integration
along the complex Bromwich contour. While the behaviour of $P(X,N)$ in 
case (A) has
been already considered in~\cite{GSPT13}, the accurate study of the
PDF in case (B) is the original result of this paper: it is in this
regime that the condensation takes place. Let us just mention that in
regime (A), which corresponds to values $X\in[-\langle X
  \rangle,\langle X \rangle]$ with $\langle X\rangle=EN$ (see inside the 
\emph{homogeneous} 
regime
($II$) in Fig.~\ref{fig:cartoon}), the PDF $P(X,N)$ exhibits a large
deviation form of the kind
\be
P(X,N) \sim \exp\left\lbrace -N \Phi\left(x=\frac{X}{N}\right) \right\rbrace,
\label{normal_ldp}
\ee
where the rate function $\Phi(x)$ was computed numerically
in~\cite{GSPT13}. It is easy to see, by virtue of central limit
theorem~\cite{Nagaev}, that in the vicinity of $X=\langle X
\rangle=EN$ and similarly around $X=-\langle X \rangle =-EN$, the rate
function simply reads

\begin{eqnarray}
\Phi(x)= \begin{cases}
\frac{(x-E)^2}{2\sigma^2}, \quad\quad\quad\quad\,\,\, {\rm for}\,\, x\lesssim E
\\
-Ex +\frac{(x+E)^2}{2\sigma^2}, \quad {\rm for}\,\, x\gtrsim -E,
\end{cases}
\label{central_rate}
\end{eqnarray}
where $x=X/N$ and $E=\langle X \rangle/N$.\\

Consider now studying $P(X,N)$ in Eq. (\ref{eq:inv-Lapl-transf}) as a
function of increasing $X$. There is a saddle point $s_0$ on the real $s$
axis as long as $-\langle X\rangle< X< \langle X\rangle$. As $X\to
\langle X\rangle$ from below, $s_0\to 0$. Similarly, as $X\to -\langle
X\rangle$ from above, $s_0\to E$ (see Fig. 3). Our main interest in
this paper is to study what happens when $X$ exceeds $\langle X\rangle
$ on the positive side (respectively when $X$ goes below $-\langle
X\rangle$ on the negative side), i.e., when there is no longer a
saddle point on the real $s$ axis in the complex $s$ plane. A detailed
study of the inverse Laplace transform in
Eq. (\ref{eq:inv-Lapl-transf}), when there is no saddle point, reveals
a rich behavior of $P(X,N)$ for $X>\langle X\rangle$ (respectively for
$X< -\langle X\rangle $).\\

\vskip 0.4cm

\noindent {\red{{\em Summary of the main results.}}}
Let us summarize our main results for $X>\langle X\rangle$ (detailed
calculations are provided in Section III). Similar computations for
$X<-\langle X\rangle$ are done in Section IV.  It turns out that when
$X$ exceeds $\langle X\rangle$ by $O(\sqrt{N})$, the behavior of
$P(X,N)$ still remains Gaussian (as expected from the central limit
theorem). Actually this Gaussian form continues to hold all the way up
to $X-\langle X\rangle \sim N^{2/3}$. However, when $X-\langle
X\rangle$ exceeds the critical value $X_c= z_c\, N^{2/3}$ (where $z_c$
is a constant of order $1$ that we compute explicitly), the Gaussian
form ceases to hold. This is where the condensate starts to form. In
this {\it intermediate} regime, where $X-\langle X\rangle = z\,
N^{2/3}$ (where $z\sim O(1)$), $P(X,N)$ exhibits an {\it anomalous}
large deviation form. Finally, in the extreme tail regime when
$X-\langle X\rangle \sim O(N)$, where the system is dominated by one
single large condensate, $P(X,N)$ has a stretched exponential
form. These three behaviors are summarized as follows:
\begin{figure}
\includegraphics[width=\columnwidth]{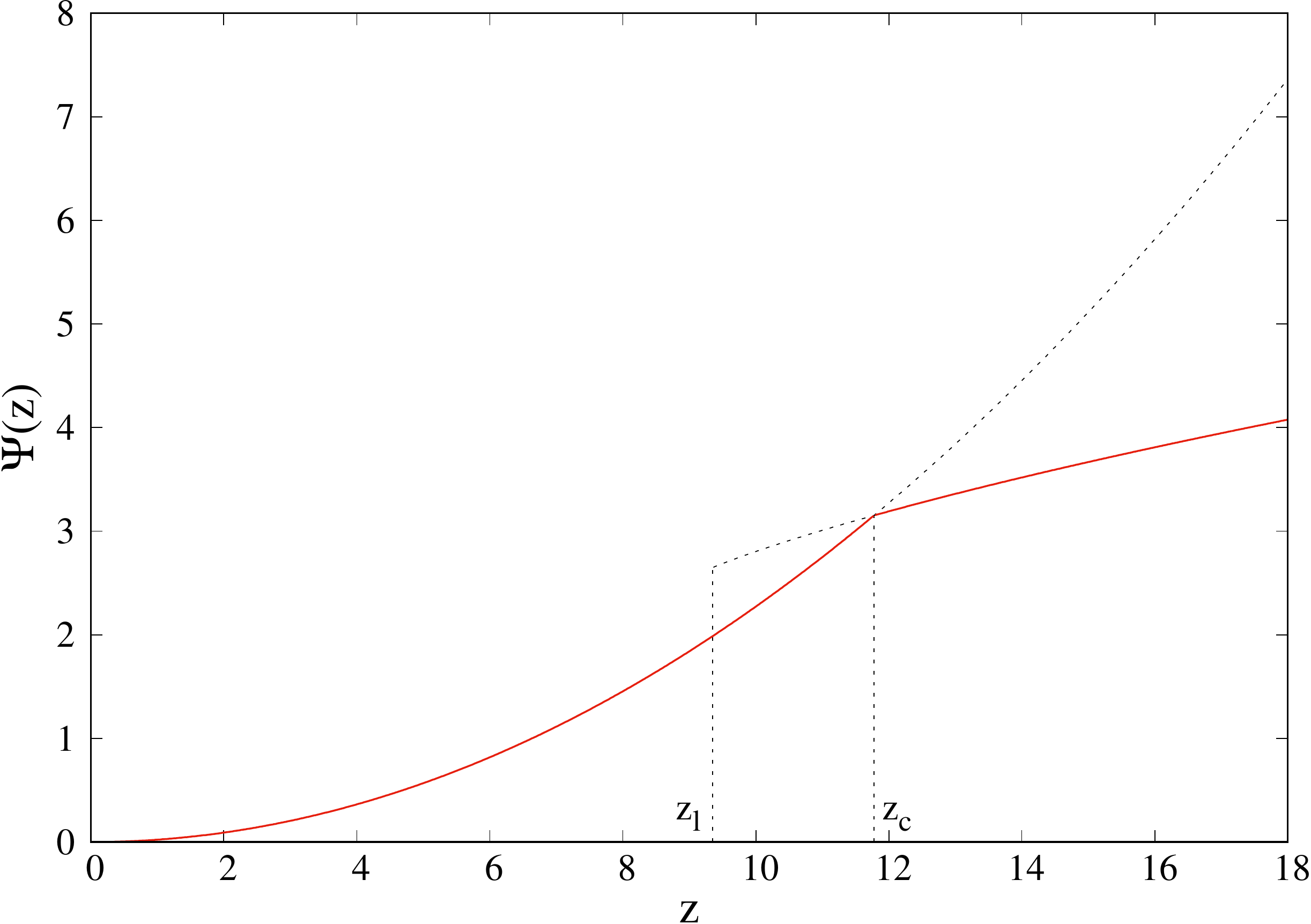}
\caption{Continuous (red) line: rate function of Eq.~(\ref{chiz.1}),
  analytical prediction. \mycol{$z_c\approx 11.78$} is the location of the
  first-order dynamical transition: $\Psi'(z)$ is clearly
  discontinuous at $z_c$. Dotted lines indicates $\chi(z)$ for $z<z_c$
  and $z^2/(2\sigma^2)$ for $z>z_c$. $z_l$ is the lowest value of $z$ such
that $\chi(z)$ can be computed via a
saddle-point approximation.}
\label{fig:transition}
\end{figure}
\bea 
P(X,N) \approx \begin{cases} e^{-(X-NE)^2/(2N\sigma^2) }  
\quad{\rm for}\quad (X - EN) \sim N^{1/2} \nonumber \\ 
e^{-N^{1/3}\Psi(z)} 
\quad\quad\quad\quad~{\rm for}\quad (X - EN)\sim N^{2/3} \nonumber \\
e^{-\sqrt{2/E}~(X-EN)^{1/2}} ~~
{\rm for}\quad (X - EN)\sim N \nonumber \\ \end{cases}
\label{three-regimes}
\eea
where $z=(X - EN)/N^{2/3}$. The rate function $\Psi(z)$ can be expressed 
as
\begin{equation}
\Psi(z)= {\rm min}\left[ \frac{z^2}{2\sigma^2}, \chi(z)\right]
\label{chiz.1}
\end{equation}
where the function $\chi(z)$ can be computed exactly (see Section III) in 
the regime
$z\in [z_l, \infty]$ with 
\begin{equation}
z_l= \frac{3}{2} \left( \frac{\sigma^4}{E} \right)^{1/3}\, .
\label{zl.1}
\end{equation}
In this regime $z_l<z<\infty$, the function $\chi(z)$ has the asymptotic 
behaviors
\be
\chi(z) = \begin{cases} 
\frac{3}{2}\left(\frac{\sigma}{E}\right)^{2/3}
\qquad\qquad\qquad\qquad z\rightarrow z_l \\ \\
\sqrt{\frac{2}{E}}
\sqrt{z}-\frac{\sigma^2}{4E}\frac{1}{z}+\mO\left(\frac{1}{z^{5/2}}\right),
\quad z\gg 1 \end{cases}
\label{eq:chi-asymptotics-intro}
\ee

The two competing functions $z^2/2\sigma^2$ and $\chi(z)$ in
Eq. (\ref{chiz.1}) are plotted
in Fig.~\ref{fig:transition}. Clearly, there exists a critical value 
$z=z_c$ 
where these
two functions cross each other, such that one gets from Eq. (\ref{chiz.1})
\be \Psi(z) = \begin{cases} z<z_c \Longrightarrow z^2/(2\sigma^2)
  \\ z>z_c \Longrightarrow \chi(z) \end{cases}.
\label{eq:rate-func}
\ee
where $z_c$ is given by the solution of the equation
\begin{equation}
\frac{z^2}{2\sigma^2}= \chi(z)
\label{eq:mcond}
\end{equation}
At $z=z_c$, the two functions match continuously, but the derivative
$\Psi'(z)$ is discontinuous at $z=z_c$ (see Fig.
(\ref{fig:transition}), signalling a first-order dynamical phase
transition. The two functions cross each other at $z_c$, provided
$z_c>z_l$. Indeed, by writing $z$ in units of $z_l$ and solving the
matching condition in Eq.~(\ref{eq:mcond}), we find that,
independently of the value of $E$,
\be 
\mycol{z_c = 2^{1/3}~z_l},
\ee
which shows that $z_c>z_l$ for any choice of the field $E$. All the
details on the computation of the function $\chi(z)$ and the
determination of $z_c$ are given in Sec.~\ref{sec:X_positive} and
in~\ref{app2}.

Our analysis also clarifies that the
mechanism of this dynamical transition is a typical one for a 
classic first-order phase transition: we show that the PDF $P(X,N)$,
for $X-\langle X\rangle= z\, N^{2/3}$ where $z\sim O(1)$, can be written 
as a
sum of two contributions, 
\be
P(X,N) = P_{\textrm{G}}(z,N) + P_{\textrm{A}}(z,N),
\ee
where $P_{\textrm{G}}(z,N)$ denotes Gaussian fluctuations, while
$P_{\textrm{A}}(z,N)$ (where the subscript $A$ is for {\it anomalous})
denotes the rare fluctuations emerging from the formation of a condensate.
These two terms compete with each other.
In the vicinity of the transition point $z_c$ both contributions can
be written in a large deviation form:
\bea
P_{\textrm{G}}(z,N) &\sim e^{-N^{1/3}z^2/(2\sigma^2)} \nonumber \\
P_{\textrm{A}}(z,N) &\sim e^{-N^{1/3}\chi(z)},
\eea
Since for $z<z_c$ one has $z^2/(2\sigma^2)<\chi(z)$, see
Fig.~\ref{fig:transition}, then
$\lim_{N\rightarrow\infty}P_{\textrm{A}}(z,N)/P_{\textrm{G}}(z,N)=0$:
the Gaussian contribution dominates. On the contrary for $z>z_c$ one
finds $z^2/(2\sigma^2)>\chi(z)$ and the probability of the condensate
takes over,
i.e. $\lim_{N\rightarrow\infty}P_{\textrm{G}}(z,N)/P_{\textrm{A}}(z,N)=0$.\\

The accurate description of the first-order dynamical phase transition
characterizing the tails of $P(X,N)$ is the main theoretical
prediction of this paper. We have verified it via direct numerical
simulations (see Fig.~\ref{fig:numerics}) and have found excellent
agreement between numerics and theory.
 
\section{First-order dynamical transition: calculation of the rate function}
\label{sec:X_positive}

In this section, we compute the large $N$ behaviour of
$P(X,N)$ for $X>\langle X
\rangle$. The strategy consists in evaluating the leading contribution
to the integral in Eq.~(\ref{eq:inv-Lapl-transf}) according to the
{\it scale} of the deviation of $X$ from the average $\langle X
\rangle$ that one is interested in. In particular we identify the
three following regimes:
\begin{enumerate}
\item $X - \langle X \rangle \sim N^{1/2}$: the \emph{Gaussian}
  regime.\\ We discuss it in Sec.~\ref{subsec:gaussian}.
\item $X - \langle X \rangle \sim N$, the \emph{extreme large-deviation}
  regime.\\ This is discussed it in Sec.~\ref{subsec:large-deviations}.
\item $X - \langle X \rangle \sim N^{2/3}$, the \emph{intermediate 
matching regime}.\\ We discuss it in Sec.~\ref{subsec:matching}.
\end{enumerate}

In the Gaussian regime, for completeness, we also repeat
how to compute $P(X,N)$ when $X<\langle X \rangle$ and $|X-\langle X
\rangle|\sim N^{1/2}$, just to show that the result is consistent with
fluctuations above the mean.\\

The three regimes listed above have one common feature: in order to
compute $P(X,N)$, the Bromwich contour appearing in its integral
representation in Eq.~(\ref{eq:inv-Lapl-transf}) must be deformed in
order to pass around the branch cut on the negative semiaxis, see
Fig.~\ref{fig:Bromwich_1}. In Fig.~\ref{fig:Bromwich_1} are
represented the analytical properties in the complex $s$ plane of
$\mL(s)$, the function defined in Eq.~(\ref{eq:inv-Lapl-transf}) and
Eq.~(\ref{eq:mL}): it has two branch cuts on the real axis. The branch
cut on the semiaxis $[E,\infty[$ is related to the behaviour of
    $P(X,N)$ for $X<-\langle X \rangle$, the branch cut on
    $]-\infty,0]$ to the behaviour for $X>\langle X \rangle$. In
Fig.~\ref{fig:Bromwich_1} are shown the examples of the two possible
shapes of the Bromwich contour, depending on whether $X$ lies inside
or outside the interval $[-\langle X \rangle,\langle X \rangle]$. For
$X\in [-\langle X \rangle,\langle X \rangle]$ the Bromwich contour is
a straight vertical line crossing the real axis at $s_0$, where $s_0$
is the saddle-point of the function $h(s)=sx+\log[\mL(s)]$, with
$x=X/N$. For $X\notin [-\langle X \rangle,\langle X \rangle]$ the
contour must be deformed in order to pass around the branch cut.

In the following subsections we discuss the details of our calculations.

\subsection{Gaussian Fluctuations}
\label{subsec:gaussian}

Let us start with the calculation of the probability of $\mO(N^{1/2})$
fluctuations around $\langle X \rangle =EN$, considering separately
the two cases $X<EN$ and $X>EN$. The result in the second case is that
the non-analiticity at the branch cut is negligible, and the
probability of fluctuations of order $|X-EN|\sim 1/\sqrt{N}$ is
Gaussian also for $X>EN$. The general strategy of all the following
calculations is to first fix the scale of the fluctuations $|X-EN|$ we
are interested in, and then consider the corresponding orders in the
expansion of $\mL(s)$ around $s_0=0$.\\

We start by computing $P(X,N)$ for $X< EN$ and $|X-EN|\sim
N^{1/2}$. The expansion of $\mL(s)$ in Eq. (\ref{eq:mL}) for small and 
\emph{positive} $s$
reads:
\be
\mL(s)=1-Es+\frac{(1+3E^2)}{2} s^2+\mO(s^3),
\ee
from which one then gets 
\be
\log[\mL(s)]=-Es+\frac{1}{2} \sigma^2 s^2+\mO(s^3),
\ee
where $\sigma^2=(2+5E^2)$ is the second cumulant of the
distribution $\mP(x)$ defined in Eq.~(\ref{x_marg.1}). Plugging the
above expansions into integral of Eq.~(\ref{eq:inv-Lapl-transf}) one
gets, for large $N$:
\be 
P(X,N) \approx \int_{s_0-i\infty}^{s_0+i\infty} \frac{ds}{2\pi i}
e^{s(X-EN)+N\frac{\sigma^2}{2}s^2+\mO(Ns^3)}.
\ee
Since we are interested in evaluating the contribution to $P(X,N)$ at
the scale $|X-EN|\sim N^{1/2}$, from $X$ and $s$ we change variables
to $z$ and $\tilde{s}$:
\bea 
X-EN &= z ~ N^{1/2} \nonumber \\
s &= \tilde{s} /N^{1/2},
\label{eq:cov-gaussian}
\eea
and then take the limit $N\rightarrow\infty$. All the irrelevant
contributions vanish and one is left with a trivial
Gaussian integral:
\be
P(X,N) = \frac{1}{\sqrt{N}}\int_{-i\infty}^{i\infty} \frac{d\tilde{s}}{2\pi i}
~ e^{\tilde{s} z+\frac{\sigma^2 }{2}\tilde{s}^2} =  \frac{e^{-(X-EN)^2/(2\sigma^2 N)}}{\sqrt{2\pi \sigma^2 N}}.
\label{eq:gaussian} 
\ee
The same result can be obtained in a straightforward manner with the
saddle-point approximation.\\

\begin{figure}
  \includegraphics[width=\columnwidth]{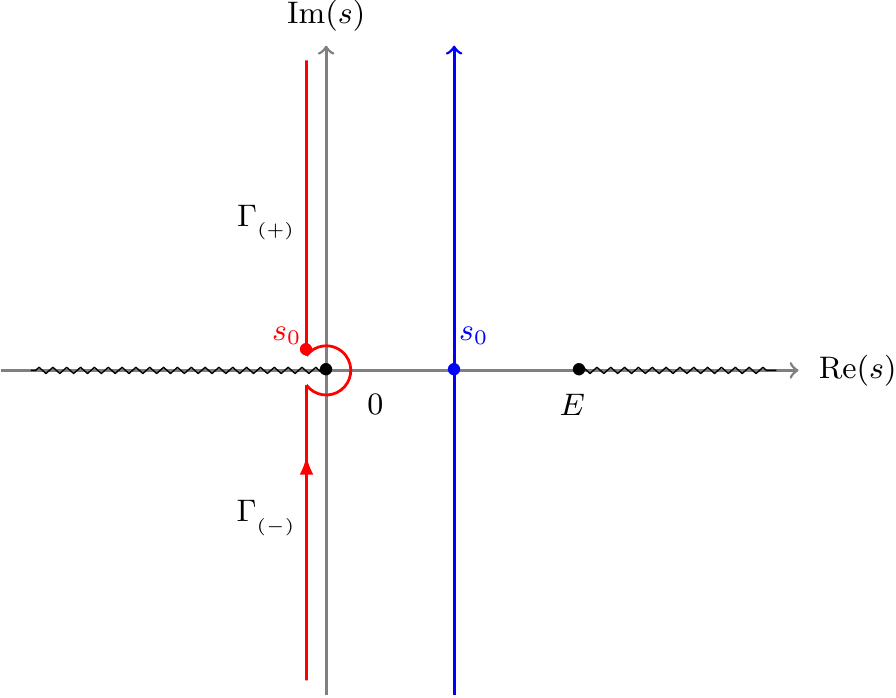}
  \caption{Analyticity structure of $\mL(s)$, see Eq.~(\ref{eq:mL}), in
  the complex $s$ plane. Wiggled lines: the two branch cuts,
  respectively $]-\infty,0]$ and $[E,\infty[$. Straight (blue) line:
      Bromwich contour for the calculation of $P(X,N)$ when $-\langle
      X \rangle <X< \langle X \rangle$, with $s_0$ indicating the
      location of the saddle-point. Deformed (red) line: Bromwich
      contour to compute $P(X,N)$ when $X>\langle X \rangle $, $s_0$
      indicates the new saddle point. $\Gamma_{(+)}$ and
      $\Gamma_{(-)}$ are labels for contour pieces in the positive and
      negative imaginary semiplanes.}
\label{fig:Bromwich_1}
\end{figure}

More interesting is the calculation of $P(X,N)$ for $X > EN$. In this
case the Bromwich contour needs to be deformed as shown in
Fig.~\ref{fig:Bromwich_1}. Due to the presence of the branch cut 
$]-\infty,0]$, the expansion of $\mL(s)$ in Eq. (\ref{eq:mL}) is 
non-analytic 
at $s_0=0$ for
    $\textrm{Re}(s)<0$, in particular it yields different results for
    the positive and the negative imaginary semiplane:
\bea 
\mL(s+i0^{+}) &= 1-Es+(1+3E^2) 
s^2+\ldots+\sqrt{\frac{2\pi}{sE}}e^{\frac{1}{2sE}+\frac{1}{2E^2}}
\nonumber \\
\mL(s+i0^{-}) &= 1-Es+(1+3E^2) 
s^2+\ldots
\nonumber \\
\label{eq:expansion-Ls}
\eea
Accordingly, for the expansion of the logarithm one finds:
\bea
\log[\mL(s+i0^{+})] &= - Es + \frac{1}{2}\sigma^2s^2 + \ldots+\sqrt{\frac{2\pi}{sE}}~e^{\frac{1}{2sE}+\frac{1}{2E^2}}\nonumber \\
\log[\mL(s+i0^{-})] &= - Es + \frac{1}{2}\sigma^2s^2 + \ldots \nonumber \\
\label{eq:expansion_s_0}
\eea

Now, to compute $P(X,N)$ at the scale $X-EN\sim\sqrt{N}$ we consider
separately the integration along the contour in the positive
immaginary semiplane, denoted as $\Gamma_{_{(+)}}$ in
Fig.~\ref{fig:Bromwich_1}, and along the contour in the negative
semiplane, denoted as $\Gamma_{_{(-)}}$, so that 
\be
P(X,N) = \mI_{(-)} + \mI_{(-)},
\ee
where the symbols $\mI_{(-)}$ and $\mI_{(+)}$ denote respectively the
contour integrations along $\Gamma_{_{(-)}}$ and $\Gamma_{_{(+)}}$. By
plugging the expansions of Eq.~(\ref{eq:expansion_s_0}) in the two
integrals and changing of variables to $z=(X-EN)/N^{1/2}$ and $s =
\tilde{s} /N^{1/2}$ one gets respectively:
\bea
\mI_{(-)} 
%%&= \int_{\Gamma_{_{(-)}}} \frac{ds}{2\pi i} e^{s z\sqrt{N}+N\frac{\sigma^2 }{2}s^2+\mO(Ns^3)} \nonumber \\
= \frac{1}{\sqrt{N}}\int_{\Gamma_{_{(-)}}} \frac{d\ts}{2\pi i} e^{\ts z+\frac{\sigma^2 }{2}\ts^2+\mO(N^{-\frac{1}{2}})} \nonumber, \\
\label{eq:Gamma_plus_gauss}
\eea
and
\bea
\mI_{(+)} 
%%&= \int_{\Gamma_{(+)}} \frac{ds}{2\pi i}
%%e^{sz\sqrt{N}+N\frac{\sigma^2 }{2}s^2+\mO(Ns^3)+N\sqrt{\frac{2\pi}{sE}}e^{\frac{1}{2sE}+\frac{1}{2E^2}}} \nonumber \\
= \frac{1}{\sqrt{N}}\int_{\Gamma_{(+)}} \frac{d\ts}{2\pi i} e^{\ts z+\frac{\sigma^2 }{2}\ts^2+\mO(N^{-\frac{1}{2}})+N^{5/4}\sqrt{\frac{2\pi}{\ts E}}~e^{\frac{\sqrt{N}}{2\ts E}+\frac{1}{2E^2}}} \nonumber, \\
\label{eq:Gamma_minus_gauss}
\eea
Since in the present case $\textrm{Re}(\ts)<0$ the non-analytic
contribution to the expansion of $\log[\mL(s)]$ in
Eq.~(\ref{eq:Gamma_minus_gauss}) is exponentially small in $\sqrt{N}$
and can be neglected, so that to the leading order the integrands of
$\mI_{(-)}$ and of $\mI_{(+)}$ are identical. By dropping also the
terms $\mO(N^{-\frac{1}{2}})$ in the argument of exponential one ends
up with the formula:
\bea
P(X,N) &= \frac{1}{\sqrt{N}}\int_{\Gamma_{_{(-)}}+\Gamma_{_{(+)}}} \frac{d\ts}{2\pi i} e^{\ts z+\frac{\sigma^2 }{2}\ts^2} = \frac{1}{\sqrt{N}}\int_{-i\infty}^{i\infty} \frac{d\ts}{2\pi i} e^{\ts z+\frac{\sigma^2 }{2}\ts^2} 
\nonumber \\
&= \frac{e^{-(X-EN)^2/(2\sigma^2 N)}}{\sqrt{2\pi \sigma^2 N}}.
\label{eq:gaussian-end}
\eea
The last equation completes the demonstration that at the scale
$|X-EN| \sim N^{1/2}$ the distribution $P(X,N)$ is a Gaussian
centered at $\langle X \rangle = EN $. This is, in fact, just a 
consequence of the validity of the central limit theorem.

\subsection{Extreme Large deviation}
\label{subsec:large-deviations}

We now focus on the extreme right tail of $P(X,N)$, where $X-NE\sim O(N)$.
To compute the leading contributions to $P(X,N)$ on this scale, we
change variables from  
from $X$ and $s$ to $z$ and $\tilde{s}$ as follows:
\bea
 X - EN &= z~N \nonumber \\
 s &= \ts/ N.
\label{eq:cov-large-deviations}
\eea
Also in this case, see for comparison Sec.~\ref{subsec:gaussian}, it
is then convenient to split the integral expression of $P(X,N)$ in the
positive and negative immaginary semiplane contributions, denoted
respectively as $\mI_{(+)}$ and $\mI_{(-)}$. The function
$\log[\mL(s)]$ is not analytic at $s_0=0$ and for $\text{Re}(s)<0$ the
expansions in the positive and negative semiplane are different and
are those written in Eq.~(\ref{eq:expansion_s_0}). Plugging in the
definition of $\mI_{(-)}$ and $\mI_{(+)}$ the expressions of
Eq.~(\ref{eq:expansion_s_0}) and the change of variables of
Eq.~(\ref{eq:cov-large-deviations}) one finds respectively
\be  \mI_{(-)} = \frac{1}{N} \int_{\Gamma_{_{(-)}}}
\frac{d\ts}{2\pi i} e^{\ts z+\frac{\sigma^2 }{2N}\ts^2+\mO(N^{-2})}
= \frac{1}{N} \int_{\Gamma_{_{(-)}}} \frac{d\ts}{2\pi
  i} e^{\ts z} \left[ 1 + \frac{\sigma^2}{2N} \ts^2 + \mO(N^{-2})
  \right],
\label{eq:Gamma_minus_large}
\ee
and
\begin{equation}
\mI_{(+)} = \frac{1}{N} 
\int_{\Gamma_{_{(+)}}} \frac{d\ts}{2\pi i} 
e^{\ts z+\frac{\sigma^2 }{2N}\ts^2+\mO(N^{-2})+
N^{3/2}\sqrt{\frac{2\pi}{\ts E}}e^{\frac{N}{2\ts E}+
\frac{1}{2E^2}}}. 
\label{eq:Gamma_plus.0}
\end{equation}
Note that all terms except $\ts z$ inside the exponential are small
for large $N$ (including the term containing $e^{N/(2\ts E)}$, since
the real part of $s$ is negative along the contour
$\Gamma_{(+)}$). Hence, we can expand the exponential for large
$N$. Keeping only leading order terms, we get
\bea \mI_{(+)} \approx \frac{1}{N} \int_{\Gamma_{_{(+)}}} \frac{d\ts}{2\pi i} e^{\ts z}
\left[ 1 + \frac{\sigma^2}{2N} \ts^2 + \mO(N^{-2}) + N^{3/2}\sqrt{\frac{2\pi}{\ts E}}e^{\frac{N}{2\ts
      E}+\frac{1}{2E^2}}\right] \nonumber \\
\label{eq:Gamma_plus_large}
\eea
Summing the two contributions and grouping the analytic terms in the expansion
one gets
\bea
\mI_{(+)} + \mI_{(-)} &\approx \frac{1}{N} \int_{-i\infty}^{i\infty}
\frac{ds}{2\pi i} e^{s z} \left[ 1 + \frac{\sigma^2}{2N} s^2 +
  \mO(N^{-2}) \right] +\nonumber \\
&+ N^{1/2}\int_{\Gamma_{(+)}} \frac{ds}{2\pi i} e^{s z }
\sqrt{\frac{2\pi}{s E}}~e^{\frac{N}{2s E}+\frac{1}{2E^2}} \, .
\label{eq:delta_plus_anomalous}
\eea
One can easily show that the integrals in the first line of
Eq. (\ref{eq:delta_plus_anomalous}) (coming from the analytic
terms) all vanish. For example, the first term just gives a
delta function $\delta(z)/N$ that vanishes for any $z>0$. 
The other analytic terms similarly can be evaluated using
\be
\int_{-i\infty}^{i\infty} \frac{d s}{2\pi i}~e^{s z}~s^n = \delta^{(n)}(z) 
\ee
and thus contribute Dirac delta's derivatives of increasing order which
all vanish for $z>0$. 

Thus, the only nonvanishing contribution for large
$N$ comes from the integral in the second line of
of Eq. (\ref{eq:delta_plus_anomalous}). To evaluate this integral,
it is convenient to first rescale $s \to \sqrt{N}\, s$ and rewrite it
as 
\be
 \mI_{(+)} + \mI_{(-)} \approx N^{3/4} \sqrt{\frac{2\pi}{E}}~e^{\frac{1}{2E^2}}  
\int_{\Gamma_{(+)}} \frac{d s}{2\pi i}~\frac{1}{\sqrt{s}}~
e^{\sqrt{N}(sz+\frac{1}{2 s E})} \, . \nonumber \\
\label{eq:Gamma_plus_large_na}
\ee
To evaluate this integral, it is first convenient to rotate
the contour $\Gamma_{(+)}$ anticlockwise by angle $\pi/2$.
We are allowed to do this since the function is analytic
in the left upper quadrant in the complex $s$ plane.
So, the deformed (rotated) contour now runs along the real axis
from $0$ to $-\infty$. This amounts to setting $s=-x$ with $x$
running from $0$ to $\infty$, and the integral in
Eq. (\ref{eq:Gamma_plus_large_na}) reduces to an integral
on the real positive axis $x\in [0, \infty]$
\begin{equation}
 \mI_{(+)} + \mI_{(-)}
\approx  N^{3/4} \sqrt{\frac{1}{2\pi\,E}}~e^{\frac{1}{2E^2}}
\int_0^{\infty} \frac{dx}{\sqrt{x}}\, e^{-\sqrt{N}\left(zx+ 
\frac{1}{2xE}\right)}\, . 
\label{wick.1}
\end{equation}
This integral can now be evaluated using the saddle point method. 
Defining,
\be
u(x) = xz + \frac{1}{2xE}\,. 
\label{eq:saddle-ux}
\ee
it is easy to check that $u(x)$ has a unique minimum at $x^*= 
1/\sqrt{2zE}$ (where $u''(x^*)>0$).
By plugging $x^*=- 1/\sqrt{2zE}$ into the integral of 
Eq.~(\ref{wick.1}) and evaluating
carefully the integral (including the Gaussian fluctuations around the
saddle point)~\cite{complex-saddle,Dennery}, we get, for large $N$ and with
$z= (X-EN)/N$,
\be
P(X,N)\approx N~e^{\frac{1}{2E^2}}
\frac{\exp\left[-(2/E)^{1/2}\sqrt{(X-EN)}\right]}{\sqrt{2E(X-EN)}},
\label{eq:large-deviations-tail}
\ee
which is our final result for this section.  Let us notice that the
expression written in Eq.~(\ref{eq:large-deviations-tail}) is
identical to $N$ times the asymptotic behaviour of the marginal 
probability
distribution of the displacement in a single run, given in
Eq.~(\ref{EP.2-text}).  This fact is perfectly consistent with the
existence of a {\em single positive condensate} for $X-NE\sim O(N)$,
that dominates the sum of $N$ i.i.d random variables each distributed via
the marginal distribution; the combinatorial factor $N$ in front indicates
that any one of the $N$ variables can be the condensate.

\subsection{First-order transition: the \emph{intermediate matching} 
regime}
\label{subsec:matching}

The main new result of this paper is the detailed study of the
intermediate regime where Gaussian fluctuation and extreme large
positive fluctuation both are of the same order: their competition is
at the heart of the first-order nature of the dynamical transition
that we find. This condition
\be
\exp\left[-\frac{(X-EN)^2}{2\sigma^2 N}\right]\sim\exp\left[-\sqrt{\frac{2}{E}}\sqrt{(X-EN)}\right],
\label{eq:matching-condition}
\ee
simply sets the scale of the {\it matching} to be
\be 
X-EN\sim N^{2/3}.
\label{eq:matching-scale}
\ee
As a consequence, in order to single out the leading contributions to
$P(X,N)$ at this scale we must change variable from $X$ to $z$ in the
integral of Eq.~(\ref{eq:inv-Lapl-transf}), with $z\sim O(1)$ such
that:
\be
X-EN = z N^{2/3}.
\ee
The {\it trick} is then to chose the proper rescaling of $s$ so that
the analytic terms of the $\log[\mL(s)]$ expansions  (responsible
for the Gaussian fluctuations), and the non-analytic ones (responsible
for the anomalous fluctuations coming from the formation of the
condensate), are of 
the same order. As is shown
below, this is achieved by rescaling $s$ as
\be s = \ts / N^{1/3}.
\ee
Once again it is useful to evaluate separately the two contributions
$\mI_{(+)}$ and $\mI_{(-)}$ after the change of variables.  Taking
into account the expansion of $\log[\mL(s)]$ in
Eq.~(\ref{eq:expansion_s_0}) one gets respectively:
\bea
\mI_{(-)} &= \int_{\Gamma_{_{(-)}}}~\frac{ds}{2\pi i}~e^{sN^{2/3}z + N\frac{\sigma^2}{2}s^2+\mO(Ns^3)} \nonumber \\
&= \frac{1}{N^{1/3}} \int_{\Gamma_{_{(-)}}}~\frac{d\ts}{2\pi i}~e^{N^{1/3}\left(\ts z + \frac{\sigma^2}{2}\ts^2\right)+\mO(1)},
\label{eq:Gamma_minus_matching}
\eea
and 
%%
%%%\begin{widetext}
\bea
\mI_{(+)} &= \int_0^{i\infty}\frac{ds}{2\pi i}~e^{s(X-NE) + N\frac{\sigma^2}{2}s^2+\mO(Ns^3)+\mycol{N} \sqrt{\frac{2\pi}{\mycol{s} E}}e^{\frac{1}{2\mycol{s} E}+\frac{1}{2E^2}}}  \nonumber \\
&= \int_0^{i\infty}\frac{d\ts}{2\pi i N^{\frac{1}{3}}}~e^{N^{1/3}\ts z + N^{1/3}\frac{\sigma^2}{2}\ts^2+\mO(1)+\mycol{N}\sqrt{\frac{2\pi N^{1/3}}{\ts E}}e^{\frac{N^{1/3}}{2\ts E}+\frac{1}{2E^2}}}  \nonumber \\
&= \int_0^{i\infty}\frac{d\ts}{2\pi i N^{\frac{1}{3}}}~e^{N^{1/3}\ts z + N^{1/3}\frac{\sigma^2}{2}\ts^2}\left[ 1 + \mycol{N} \sqrt{\frac{2\pi N^{1/3}}{\ts E}}e^{\frac{N^{1/3}}{2\ts E}+\frac{1}{2E^2}} \right]. \nonumber \\
\label{eq:Gamma_plus_matching}
\eea
%%%\end{widetext}
%%
By summing the expression of the two integrals in
Eq.~(\ref{eq:Gamma_minus_matching}) and
Eq.~(\ref{eq:Gamma_plus_matching}) it is then easy to write
$P(X,N)$, with $X-E\,N= z\, N^{2/3}$, explicitly as the sum of a Gaussian 
and an anomalous
contribution:
\bea
P(X,N) &= P_{\textrm{G}}(z,N) + P_{\textrm{A}}(z,N) \nonumber \\
P_{\textrm{G}}(z,N) &= \frac{1}{N^{1/3}} 
\int_{-i\infty}^{i\infty}~\frac{ds}{2\pi i}~e^{N^{1/3}\left(s z + \frac{\sigma^2}{2}s^2\right)} \nonumber \\
P_{\textrm{A}}(z,N) &= N^{5/6}~\frac{e^{1/(2E^2)}}{i\sqrt{2\pi E}}~\int_{\Gamma_{_{(+)}}}ds\frac{1}{\sqrt{s}}~e^{N^{1/3}F_z(s)} \nonumber \\
\label{eq:PX-matching}
\eea
where the function $F_z(s)$ reads:
\be
F_z(s) = sz + \frac{1}{2}\sigma^2s^2 +\frac{1}{2sE}.
\label{eq:Fz-matching}
\ee
\begin{figure}
  \includegraphics[width=\columnwidth]{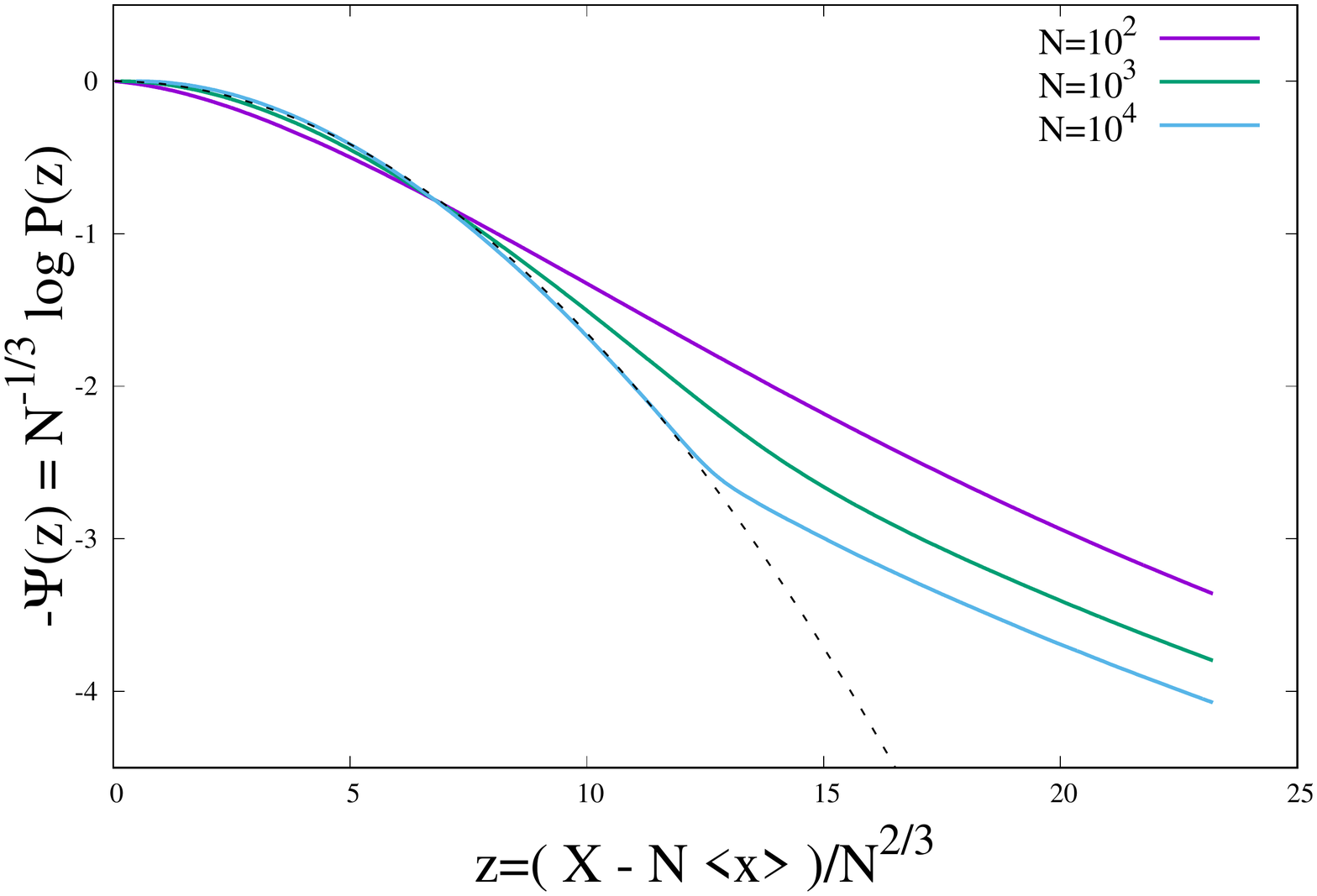}
  \includegraphics[width=\columnwidth]{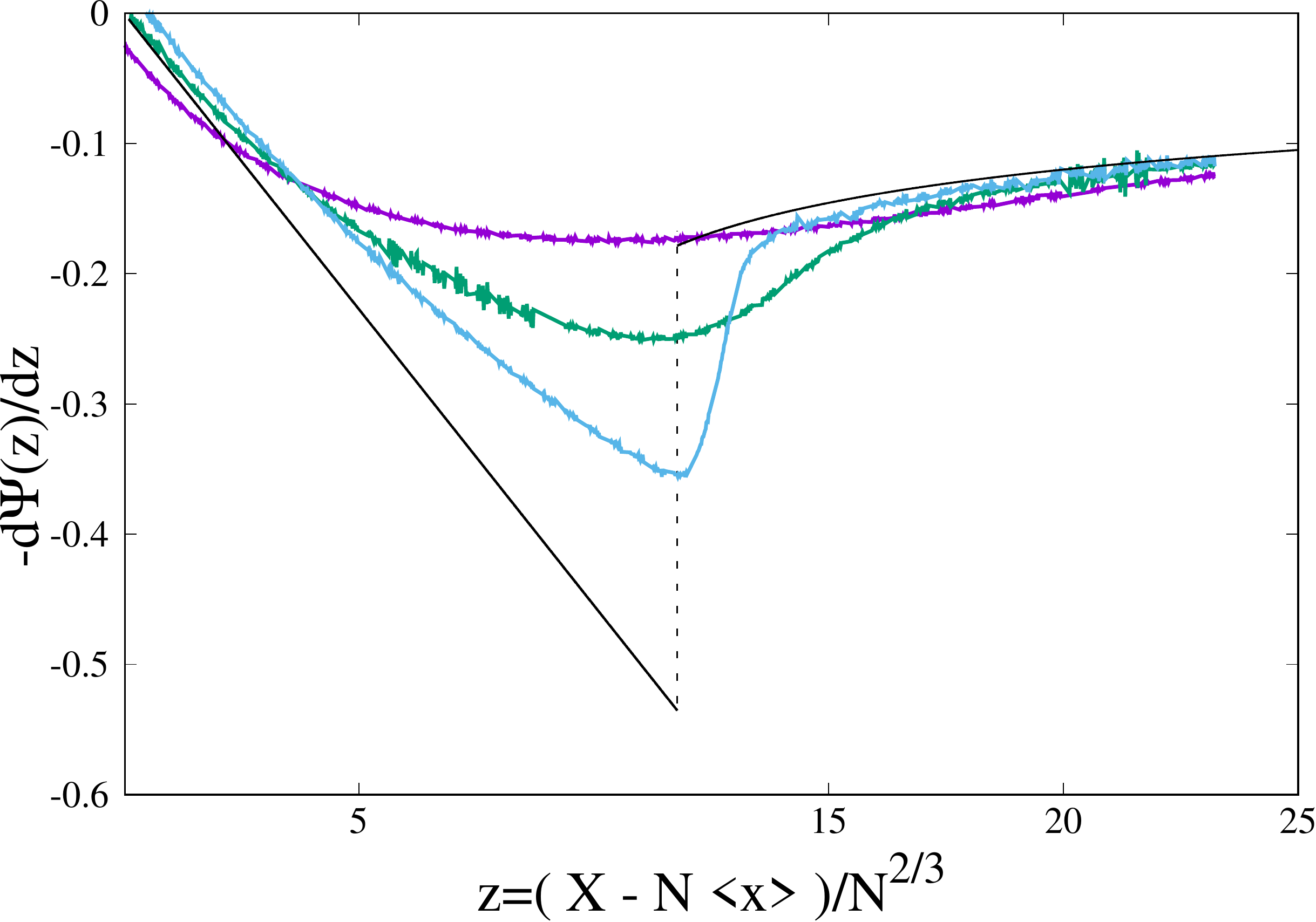}
\caption{ \emph{Top}: numerical data for the rate function
  $\Psi(z)$. Different curves correspond to different number of runs
  $N$ in the trajectory: $N=10^2, 10^3, 10^4$. Acceleration is set to
  $E=2$. Dotted (black) line: guide to the eye, Gaussian (inverted)
  parabola. \emph{Bottom}: numerical data for the rate function
  derivative $\Psi'(z)$. Continuous black line is the analytical
  prediction in the limit $N\rightarrow\infty$, the coordinate of the
  transition point is \mycol{$z_c\approx 11.78$} (for $E=2$).}
\label{fig:numerics}
\end{figure}

The integral in the second line of Eq.~(\ref{eq:PX-matching}) can be
easily performed using the saddle point method, and gives a Gaussian
contribution, justifying its name
\begin{equation}
P_{\textrm{G}}(z,N)\approx \frac{1}{\sqrt{2\pi \sigma^2 N}}\,\exp\left[- 
N^{1/3}\, 
\frac{z^2}{2\sigma^2}\right]\, .
\label{PGXN.1}
\end{equation}
The integral in the second line of Eq.~(\ref{eq:PX-matching}), giving
rise to the anomalous part, can 
be also be computed with a saddle point approximation only when the
saddle-point equation
\be \frac{\partial F_z(s)}{\partial s}=0
\label{eq:spe-F}
\ee
has a real root $s^*$. The real roots of Eq.~(\ref{eq:spe-F}) and
their properties are studied in full detail in~\ref{app2}. In the same
appendix, we also discuss the domain of existence of the saddle point
solution and give the explicit solution.

Skipping further details, we find that the saddle-point equation
$F'_z(s)=0$ has a real solution $s^*$ only for $z\in [z_l,\infty]$,
with $z_l=3\sigma^{4/3}/{2E^{1/3}}$ (see~\ref{app2}): in this range
the integral $P_{\textrm{A}}(z,N)$ can be explicitly evaluated. For
$z<z_l$, computing the integral is hard as there is no saddle point on
the real negative $s$ axis. However, as we will see, we do not need
the information on $P_{\textrm{A}}(z,N)$ for $z=(X-EN)/N^{2/3}<z_l$.
We will see that the transition occurs at $z=z_c>z_l$, so it is enough
to compute $P_{\textrm{A}}(z,N)$ for $z>z_l$. Hence, for our purpose,
evaluating $P_{\textrm{A}}(z,N)$ by saddle point is
sufficient. Assuming the existence of a saddle point at $s=s^*$ and
plugging the explicit expression of $s^*$ as a function of $z$
(see~\ref{app2}) into Eq.~(\ref{eq:PX-matching}) one gets:
\be
P_{\textrm{A}}(z,N) \sim~e^{-N^{1/3}\chi(z)}.
\label{eq:anomalous-int}
\ee
The shape of the function $\chi(z)$ is shown in
Fig.~\ref{fig:transition}, where its behavior is compared with the
parabola $z^2/(2\sigma^2)$ of the Gaussian term. All the details on
the derivation of $\chi(z)$ are in~\ref{app2}. The asymptotics are:
\be
\chi(z) = \begin{cases} \frac{3}{2}\left(\frac{\sigma}{E}\right)^{2/3}\qquad\qquad\qquad\qquad z\rightarrow z_l \\ \\
\sqrt{\frac{2}{E}}
\sqrt{z}-\frac{\sigma^2}{4E}\frac{1}{z}+\mO\left(\frac{1}{z^{5/2}}\right)
\quad z\gg 1 \end{cases}
\label{eq:chi-asymptotics}
\ee
Dropping the irrelevant prefactors aside, we then get 
\be P(X,N) \approx \exp\left\lbrace
-N^{1/3}\frac{z^2}{2\sigma^2}\right\rbrace+\exp\left\lbrace
-N^{1/3}\chi(z)\right\rbrace.
\ee
The equation $z^2/(2\sigma^2)=\chi(z)$ can be solved exactly with a simple
argument (see~\ref{app2:zc}), yielding the following value of $z_c$ in
units of $z_l$:
\be
\mycol{\frac{z_c}{z_l} = 2^{1/3}}
\label{eq:zc_unit_zl}
\ee
In the numerical simulations we set the acceleration to $E=2$.  By
plugging this value in the expression of $z_l$ given in~\ref{app2} we
finally get:
\be
\mycol{z_c \approx 11.78},
\label{eq:zc_abs_E2}
\ee
the value indicated by the dotted vertical line in Fig.~\ref{fig:numerics}.
\\

The mechanism of the first-order transition is now transparent.
Recall that $X-EN= z\, N^{2/3}$. When
$z<z_c$, the probability distribution $P(X,N)$ is dominated by the
Gaussian contribution $P_{\textrm{G}}(z,N)$, since 
$z^2/(2\sigma^2)<\chi(z)$. On
the contrary for $z>z_c$ the distribution is dominated by the
\emph{anomalous} contribution $P_{\textrm{A}}(z,N)$, since
$z^2/(2\sigma^2)>\chi(z)$.  The result can be summarized as follows:
\bea
z<z_c &\Longrightarrow \frac{z^2}{2\sigma^2} < \chi(z) \Longrightarrow P(X,N)\approx e^{-N^{1/3}z^2/(2\sigma^2)} \nonumber \\
z>z_c &\Longrightarrow \frac{z^2}{2\sigma^2} > \chi(z) \Longrightarrow P(X,N)\approx e^{-N^{1/3}\chi(z)}
\label{eq:summary-transition}
\eea
Thus the mechanism behind the first order transition corresponds to a
classic first-order phase transition scenario in standard
thermodynamics: in the vicinity of the transition point there is a
competition between two phases characterized by a different
free-energy (here the value of the rate function) and the transition
point itself is defined as the value of the control parameter (here
the value of the displacement) where the free-energy difference
between the two phases changes sign.\\

\subsection{First-order transition from numerical simulations}
\label{subsec:simulations}

{\red{ The direct consequence of the description in
Eq.~(\ref{eq:summary-transition}) is that the rate function $\Psi(z)$,
which is $\Psi(z)=z^2/(2\sigma^2)$ for $z<z_c$ and $\Psi(z)=\chi(z)$
for $z>z_c$, has a discontinuity in its first-order derivative
$\Psi'(z)$ at $z_c$: this happens because the two functions
$z^2/(2\sigma^2)$ and $\chi(z)$ match continuously at $z_c$, but with
a different slope.\\

We show in this section that the discontinuity of the rate function
$\Psi(z)$ appears for large enough $N$ when one tries to sample
numerically the tails of $P(X,N)$. We have studied the behaviour of
$P(X,N)$ for $N=10^2,10^3,10^4$ runs in the trajectory. The behaviour
of the rate function $\Psi(z)$ and of its derivative $\Psi'(z)$ are
shown in Fig.~\ref{fig:numerics}. While at $N=10^2$ the transition
from the Gaussian to the large deviations regime is still a smooth
crossover, the trend for increasing $N$ goes clearly towards a
discontinuous jump of $\Psi'(x)$. The location of the discontinuity
revealed by the numerical simulations is in agreement with the
analytic prediction \mycol{$z_c\approx 11.78$} given for the value $E=2$.\\

Simulations are straightforward but one has to chose a clever
strategy: just looking at the probability distribution of independent
identically distributed random variables is not sufficient, since
doing like that one can only probe the {\it typical} fluctuations
regime, $|X-\langle X \rangle| \sim N^{1/2}$, but not the large
deviations. In order to sample $P(X,N)$ in the whole regime of
interest a set of many simulations is needed, each probing the
behaviour of the PDF in a narrow interval $[X^*,X^*+\Delta]$. We
provide in what follows a detailed description of the numerical
protocol.\\

In order to achieve an efficient sampling of $P(X,N)$ also in the {\it
  matching} and in the {\it large deviations} regime we follow here
the strategy to compute the tails of random matrices eigenvalues
distribution used in~\cite{NMV10,NMV11}. The basic idea is to sample
$P(X,N)$ in many small intervals $[X^*,X^*+\Delta]$ varying $X^*$, in
order to recover finally the whole distribution. The sampling of
$P(X,N)$ in each interval $[X^*,X^*+\Delta]$ corresponds to an
independent Monte Carlo simulation. Since the total number of runs in
the trajectory is fixed to $N$, for each value of $X^*$ we fix the
initial condition chosing a set
$(\tau_1^{\text{in}},\ldots,\tau_N^{\text{in}})$ of runs durations
and a set $(v_1^{\text{in}},\ldots,v_N^{\text{in}})$ of initial
velocities for each run such that:
\be
\sum_{i=1}^N \left[ v_i^{\text{in}}\tau_i^{\text{in}}+\frac{E}{2}(\tau_i^{\text{in}})^2 \right] > X^*.
\ee
In the intial condition all the local variables are of order unit:
$v_i^{\text{in}},\tau_i^{\text{in}} = \mO(1)$. The stochastic dynamics to sample
$P(X,N)$ in the vicinity of $X^*$ then goes on as any standard
Metropolis algorithm: attempted updates are accepted or rejected with
probability $p=\text{min}[1,p(\mC^{\text{old}})/p(\mC^{\text{new}})]$
where the stationary probability of a configuration $p(\mC)$ reads as:
\be
p(\tau_1,v_1,\ldots,\tau_N,v_N) \approx \exp\left(-\sum_{i=1}^N \left[\tau_i+v_i^2/2\right]\right). 
\ee
The only additional ingredient with respect to a standard Metropolis
algorithm is that all attempts which brings $X_N$ below $X^*$ are
rejected.  The precise form of the probability distribution sampled
for each value of $X^*$ in the Monte Carlo dynamics is therefore:
\be
P(X,N | X>X^*) = e^{-\sum_{i=1}^N [\tau_i+v_i^2/2])}~\Theta\left( \sum_{i=1}^N \left[ v_i\tau_i + \frac{E}{2}\tau_i\right] - X^*\right),
\label{eq:prob-biased}
\ee
where $\Theta(x)$ is the Heavyside step function.\\ We define as a
Monte Carlo {\it sweep} the sequence of $N$ local attempts of the kind
$(v_i,\tau_i)~\Rightarrow~(v_i^{\text{new}},\tau_i^{\text{new}})$ where
\be
v_i^{\text{new}} = v_i + \delta v, \nonumber 
\ee
\be
\tau_i^{\text{new}} = \tau_i + \delta \tau.
\ee
The shifts $\delta v$, $\delta \tau$ are random variables drawn from 
uniform distributions. Then, {\it if and only if} the constraint
implemented by the Heavyside step function in
Eq.~(\ref{eq:prob-biased}) is satisfied, otherwise the attempt is
rejected immediately, the new values
$(v_i^{\text{new}},\tau_i^{\text{new}})$ are accepted with
probability:
\be
p_{\text{acc}} = \text{min}\left[ 1,e^{-\left[(v_i^{\text{new}})^2/2+\tau_i^{\text{new}}-(v_i^2/2+\tau_i)\right]}\right].
\ee
In order to recover the full probability distribution $P(X,N)$ within
the interval $X\in \mathcal{I}=[\langle X \rangle,\langle X \rangle +
  2 z_c \langle X \rangle^{2/3}]$ we have divided $\mathcal{I}$ in a
grid of $M=100$ elements.  More precisely, we have run $M$
simulations for a set of equally-spaced values $X^*\in \mI$. Each
simulation allows one to sample $P(X,N | X>X^*)$ only in a small
interval on the right of $X^*$. The value $z_c$ is the critical one
predicted by the theory. The choice of the interval $\mI$ has been
somehow arbitrary, we just took care that it was centered around the
expected critical value $X_c$ for the condensation transition. We have
taken a number of sweeps $N_{\text{sweeps}}\approx 10^7$, enough to
forget the initial conditions.

The relation between the PDF $P(X,N | X>X^*)$ sampled in the MC
numerical simulations and the PDF we want to investigate, $P(X,N)$, is
as follows:
\be 
P(X,N) = P(X,N|X>X^*)~P(X>X^*,N)
\label{eq:product}
\ee
In particular, what we are interested in is the rate function $\Psi(z)$ defined as:
\be
\Psi\left(z=\frac{X-\langle X \rangle}{N^{2/3}}\right) = -\frac{1}{N^{1/3}}~\log\left[P(X,N)\right]. 
\ee
The rate function $\Psi_{z^*}(z)$ that we measure in the vicinity of
$X^*$ by sampling the probability distribution in
Eq.~(\ref{eq:prob-biased}) differs from the original one, due to
Eq.~(\ref{eq:product}), by an additive constant:
\bea 
\Psi(z) &= -\frac{1}{N^{1/3}} \log\left[ P(X,N|X>X^*) \right] -\frac{1}{N^{1/3}} \log\left[ P(X>X^*,N) \right]  \nonumber \\
&= \Psi_{z^*}(z) + f(z^*), \nonumber \\
\eea
where $f(z^*)$ is a function that depends only on $z^*$. By taking the
derivative with respect to $z$ (and taking into account that $dz =
dX/N^{2/3}$) one gets rid of the additive constant and obtain the
following expression:
\be 
\frac{d\Psi(z)}{dz} = \frac{d\Psi_{z^*}(z)}{dz}  = -\frac{N^{1/3}}{P(X,N|X>X^*)} \frac{d P(X,N|X>X^*) }{dX} 
\ee
Therefore what we have done has been to sample numerically
$d\Psi_{z^*}(z)/dz$ in the vicinity of many values $z^*$ by means of
the biased Monte Carlo dynamics. The function $\Psi(z)$ has been then
obtained from the numerical integration of the first-order
derivative. Both the rate function $\Psi(z)$ and its first derivative
$\Psi'(z)$ are shown in Fig.~\ref{fig:numerics}.}} \\ \\ 

\section{First-Order transition for \emph{negative} fluctuations.}
\label{sec:X_negative}

So far we focused only on the right tail of $P(X,N)$, but the
probability distribution is not symmetric due to $E$, as is shown in
the pictorial representation of Fig.~\ref{fig:cartoon}. We need to
comment also on the behaviour of the left tail of $P(X,N)$. In this
section we demonstrate that even for negative fluctuations a
first-order dynamical transition takes place, and that, following the
same arguments of Sec.~\ref{sec:X_positive}, it is located at
$X_c^{(-)}=-X_c$, where $X_c=z_c N^{2/3}+NE$, with $z_c$ given in
Eqs.~(\ref{eq:zc_unit_zl}) and (\ref{eq:zc_abs_E2}). The location of
the transition for negative fluctuations is symmetric to that for
positive ones. The only difference with the case of positive
fluctuations is that for $X<0$ the PDF has in front an exponential
damping factor due to the field $E$. Here is the summary of the
behavior of $P(X,N)$ in the three regimes, i.e. typical fluctuations,
extreme large negative deviations and the intermediate matching
regime, for $X<0$:
%%
%\begin{widetext}
\bea
P(X,N) \approx \begin{cases} e^{-E|X|}~e^{-(X+NE)^2/(2N\sigma^2) }  ~~{\rm for}~(X + EN) \sim N^{1/2} \nonumber \\ 
e^{-E|X|}~e^{-N^{1/3}\Psi(z)} \quad\quad\quad\quad{\rm for}~(X + EN)\sim N^{2/3} \nonumber \\
e^{-E|X|}~e^{-\sqrt{2/E}~|X+EN|^{1/2}} ~~{\rm for}~(X + EN)\sim N \end{cases}, \nonumber \\
\label{three-regimes-negative}
\eea
%\end{widetext}
%%
where $z=-(X+NE)/N^{2/3}$ and the rate function $\Psi(z)$ is the same
as te one we have computed for positive fluctuations.\\

The calculations to obtain the behaviours in
Eq.~(\ref{three-regimes-negative}) are identical to those for $X>0$,
which are described in full detail in Sec.~(\ref{sec:X_positive}) and
in~\ref{app2}. We are not going to repeat all of them here. We will
sketch the derivation of the results in
Eq.~(\ref{three-regimes-negative}) only for the {\it matching regime},
which is the most interesting among the three. This discussion has
also the purpose to highlight the (small) differences with the
calculations in the case $X>0$, in particular to show where the
prefactor $e^{-E|X|}$ comes from.\\

First, as can be easily noticed looking at Eq.~(\ref{eq:mL}),
the function $\mL(s)$ has the following symmetry
\be
\mL(s) = \mL(E-s),
\label{eq:symmetry-mL}
\ee
which is important for the following reason. To compute $P(X,N)$ for
values of the total displacemente $X> EN$ we needed to wrap the
Bromwich contour around the branch cut at $]-\infty,0]$ (see
    Fig.~\ref{fig:Bromwich_1}). This was done by taking the analytic
    continuation of $\log[\mL(s)]$ in the complex plane in the
    neighbourhood of $s_0=0$. In the same manner, in order to compute
    $P(X,N)$ for $X<-NE$, we must wrap the Bromwich contour around the
    branch cut at $[E,\infty[$. To do this we need the analytic
        continuation of $\log[\mL(s)]$ in the neighbourhood of
        $s_0=E$. Due to the symmetry in Eq.~(\ref{eq:symmetry-mL}) the
        expansion of $\log[\mL(s)]$ in the neighbourhood of $s_0=E$ is
        identical to the one in the neighbourhood of $s_0=0$,
        including the non-analyticities due to the branch cut. In
        particular we have that:
\bea
 \log[\mL(E-s+i0^{+})] &= - E (E-s) + \frac{1}{2}\sigma^2(E-s)^2 + \ldots+\sqrt{\frac{\pi}{2E(E-s)}}e^{\frac{1}{2s(E-s)}+\frac{1}{2E^2}}\nonumber \\
 \log[\mL(E-s+i0^{-})] &= - E (E-s) + \frac{1}{2}\sigma^2(E-s)^2 + \ldots
\label{eq:expansion_s_E}
\eea
As done in the case of positive fluctuations, also for $X<-NE$ is
convenient to split the expression of the inverse Laplace transform of
$P(X,N)$, see Eq.~(\ref{eq:inv-Lapl-transf}), in two contributions:
the contour integral in the negative semiplane, $\mI_{(-)}$, and the
contour integral in the positive semiplane, $\mI_{(+)}$.  Let us
consider first the integral $\mI_{(-)}$:
\be
\mI_{(-)} = \int_{\Gamma_{(-)}} \frac{ds}{2\pi i} e^{sX + N \left[ - E (E-s) + \frac{1}{2}\sigma^2(E-s)^2 +\ldots \right]}.
\ee
In this case ($X<0$) is convenient to change variable from $s$ to
$y=E-s$:
\bea
 \mI_{(-)} &= \int_{-\Gamma_{(-)}} \frac{dy}{2\pi i}~e^{(E-y)X + N \left[ - E y + \frac{1}{2}\sigma^2y^2 +\ldots \right]} \nonumber \\
 &= e^{XE} ~\int_{\Gamma_{(-)}} \frac{dy}{2\pi i}~e^{-y(X+EN) + N \frac{1}{2}\sigma^2y^2 +\ldots}. 
\eea
Then, in order to have a variable which is positive and is of order
$\mO(1)$ when $X+EN\sim N^{2/3}$ we introduce:
\be
z = - \frac{X+EN}{N^{2/3}}. 
\ee
By rescaling the integration variable 
$y=\tilde{y}/N^{\frac{1}{3}}$, appropriate for the matching 
intermediate regime, we can rewrite
\be
\mI_{(-)} = e^{EX} ~\frac{1}{N^{1/3}}\int_{\Gamma_{(-)}}\frac{d\tilde{y}}{2\pi i}~e^{N^{1/3}\left[\tilde{y}z + \frac{1}{2}\sigma^2\tilde{y}^2\right]+\mO(1)}.
\label{eq:Iminus_Emeno}
\ee
The expression of $\mI_{(-)}$ in Eq.~(\ref{eq:Iminus_Emeno}), is,
apart from the prefactor $e^{EX}$, identical to the analogous one
evaluted for $X>0$, see Eq.~(\ref{eq:Gamma_minus_matching}). The only
difference is that now the scaling variable $z$ is defined as $z =
-(X+EN)/N^{2/3}$ rather than $z=(X-EN)/N^{2/3}$.  In the same way for
the integral in the positive complex semiplane we find:
%%
%\begin{widetext}
  \bea \mI_{(+)}= e^{EX}\frac{1}{N^{1/3}}\int_{\Gamma_{_{(+)}}}
  \frac{d\tilde{y}}{2\pi i}
  e^{N^{1/3}\left[\tilde{y}z+\frac{1}{2}\sigma^2 \tilde{y}^2\right]+
    \ldots+N\sqrt{\frac{2\pi N^{1/3}}{E\tilde{y}}}
    e^{\frac{N^{1/3}}{2E\tilde{y}}+\frac{1}{2E^2}}}
  \label{eq:Gamma_plus_minus}
  \eea
Recalling that we are expanding for $\text{Re}(y) =
\text{Re}(E-s) <0$ (see Fig.\ref{fig:Bromwich_1}), and 
hence $\tilde{y}<0$, we can further expand:
\bea
\mI_{(+)}&=& e^{EX}\frac{1}{N^{1/3}}\int_{\Gamma_{_{(+)}}} \frac{d\tilde{y}}{2\pi i}  e^{N^{1/3}\left[\tilde{y}z+\frac{1}{2}\sigma^2 \tilde{y}^2\right]}\left[ 1+N \sqrt{\frac{2\pi N^{1/3}}{E\tilde{y}}} e^{\frac{N^{1/3}}{2E\tilde{y}}+\frac{1}{2E^2}} \right] \nonumber \\
&=& e^{EX}\frac{1}{N^{1/3}}\int_{\Gamma_{_{(+)}}} \frac{d\tilde{y}}{2\pi i}  e^{N^{1/3}\left[\tilde{y}z+\frac{1}{2}\sigma^2 \tilde{y}^2\right]} + N^{5/6}~e^{EX}~
\frac{e^{1/(2E^2)}}{i\sqrt{2\pi E}}\int_{\Gamma_{_{(+)}}} \frac{d\tilde{y}}{\sqrt{\tilde{y}}}~e^{N^{1/3}F_z(\tilde{y})} \nonumber \\
\eea
so that
\bea
 \mI_{(+)} + \mI_{(-)} &= e^{EX}\frac{1}{N^{1/3}}\int_{-i\infty}^{i\infty} \frac{d\tilde{y}}{2\pi i} e^{N^{1/3}\left[\tilde{y}z+\frac{1}{2}\sigma^2 \tilde{y}^2\right]} + N^{5/6}~e^{EX}~\frac{e^{1/(2E^2)}}{i\sqrt{2\pi E}} \int_{\Gamma_{_{(+)}}} \frac{d\tilde{y}}{\sqrt{\tilde{y}}}  e^{N^{1/3}F_z(\tilde{y})} \nonumber \\
F_z(\tilde{y}) &= \tilde{y}z + \frac{1}{2}\sigma^2\tilde{y}^2 +\frac{1}{2\tilde{y}E},
\label{eq:negative-matching}
\eea
%%
%\end{widetext}
where the function $F_z(\tilde{y})$ is identical to that of
Eq.~(\ref{eq:Fz-matching}), hence leading to the same conclusions.
For negative fluctuations as well, it is then straightforward to see that 
in the
intermediate matching regime we have two competing contributions, i.e. the
Gaussian, $P_{\textrm{G}}(z,N)$, and anomalous one, $P_{\textrm{A}}(z,N)$:
\bea
P_{\textrm{G}}(z,N) = \frac{1}{N^{1/3}}\int_{-i\infty}^{i\infty} 
\frac{d\tilde{y}}{2\pi i} e^{N^{1/3}\left[\tilde{y}z+\frac{1}{2}\sigma^2 \tilde{y}^2\right]} \nonumber \\
P_{\textrm{A}}(z,N) = N^{5/6}~e^{EX}~\frac{e^{1/(2E^2)}}{i\sqrt{2\pi E}}\int_{-i\infty}^{i\infty} 
\frac{d\tilde{y}}{\sqrt{\tilde{y}}}~e^{N^{1/3}F_z(\tilde{y})},
\eea
where $z= - (X+NE)/N^{2/3}$. The probability distribution for
negative fluctuations in the matching regime reads therefore as:
\be
P(X,N) = e^{EX} \left[ P_{\textrm{G}}(z,N) + P_{\textrm{A}}(z,N) \right],
\ee
Apart from the prefactor $e^{EX}$ the expression of $P(X,N)$, for
negative fluctuations in the matching regime, is the same as that for
positive fluctuations: the condensation transition at $z_c$ is driven
by the same mechanism, the competition between the Gaussian
fluctuations of $P_{\textrm{G}}(z,N)$ and the anomalous one of
$P_{\textrm{A}}(z,N)$.
The calculation of the probability of typical fluctuations $X+NE \sim
N^{1/2}$ and of large deviations $X+NE\sim N$ can be very easily done
following the same steps of Sec.~\ref{sec:X_positive}, which we do not
repeat here.

\section{Conclusions}
\label{sec:conclusions}

{\red{We have studied the probability distribution $P(X,N)$ of the total
displacement $X_N=\sum_{i=1}^N x_i$ for a Run-and-Tumble (RTP) particle on a
line, subject to a constant force $E>0$. The PDF $p(\tau)$ for the
distribution of duration of a run and the PDF $q(v)$ for the velocity at the
begining of a run are two inputs to the model, along with $E>0$.
The standard RTP corresponds to the choice $E=0$, $p(\tau)=e^{-\tau}\, \theta(\tau)$ (Poisson tumbling with rate $1$)
and a bimodal $q(v)= \frac{1}{2}\left[\delta(v-v_0)+\delta(v+v_0)\right]$.
The main conclusion of this paper is that a broad class of $p(\tau)$ and $q(v)$, for $E>0$, leads
to a condensation transition. This is manifest as a singularity in the displacement PDF $P(X,N)$
for large $N$ and the transition is first-order. A criterion for the condensation is provided
for different choices of $p(\tau)$ and $q(v)$. As a representative case, we have provided detailed
analysis and results for the specific choice: arbitrary $E>0$, $p(\tau)= e^{-\tau}\, \theta(\tau)$ and 
$q(v)= e^{-v^2/2}/\sqrt{2\pi}$. We have also argued that the standard RTP does not have this
interesting phase transition.

By a detailed computation of the PDF $P(X,N)$ of the total displacement after $N$ runs,}}
we have shown that while the central part of the PDF $P(X,N)$ is characterized
by a Gaussian form (as dictated by the central limit theorem), both
the right and left tails of $P(X,N)$ have anomalous large deviations
forms. On the positive side, as the control parameter $X-EN$ exceeds a
critical value $z_c\, N^{2/3}$, a condensate forms, i.e, the sum
starts getting dominated by a single long run. This signals a phase
transition, as a function of $X$, from the central regime dominated by
Gaussian fluctuations to the condensate regime dominated by a single
long run. A similar transition occurs for large negative $X$ where a
negative long run dominates the sum. The phase transition is
qualitatively similar to condensation phenomenon in mass transport
models, where the role of the large condensate mass is played here by
the macroscopic extent of the displacement travelled without tumbles
in one single run.

The main new result of our study is the uncovering of an intermediate
matching regime where the PDF $P(X,N)$ of the total displacement
exhibits an anomalous large deviation form, $P(X,N)\sim
e^{-N^{1/3}\Psi(z)}$ with $z=(X-\langle X\rangle)/N^{2/3}$.  Quite
remarkable is the non-analytic behaviour of the associated rate
function $\Psi(z)$ at the critical point $z=z_c$, here the function is
continuous but its first derivative jumps: we are in presence of a
first-order phase transition. The two phases on either side of the
critical point $z_c$ corresponds respectively to a fluid phase
($z<z_c$) and a phase with a single large condensate ($z>z_c$).  The
mechanism behind this transition is typical of a thermodynamic
first-order phase transition, where there is an energy jump (first
order derivative of the free energy with respect to the inverse
temperature $\beta$) emerging from the competition between two phases.
Here we have homogeneous trajectories with Gaussian probability
$P_{\textrm{G}}(X,N)$ competing with trajectories dominated by one 
single run characterized by the anomalous part of the distribution 
$P_{\textrm{A}}(X,N)$. The transition takes place
when the two competing terms are of the same order. An interesting
feature of the analysis presented here is that the first-order
dynamical transition studied takes place in a regime where the natural
scale (speed) of large deviations is $N^{1/3}$ and not $N$, as is
typical in extensive thermodynamic systems.\\

In this paper, we have shown that the problem of computing the total
displacement of the RTP reduces to the computation of the distribution
of the linear statistics (in this case just the sum) of a set of i.i.d random variables, each
drawn from a marginal distribution that has a stretched exponential
tail. Our study shows that even for such a simple system, the
distribution $P(X,N)$ has an anomalous large deviation regime that
exhibits a discontinuity in the first-derivative of the rate
function. It is worth pointing out that in a certain class of strongly
correlated random variables (typically arising in problems involving
the eigenvalues of a random matrix), the distribution of linear
statistics is known to exhibit a large deviation form that typically
undergoes similar phase
transitions~\cite{VMB08,MNSV09,VMB10,NMV10,NMV11,MNSV11,TM13,MS14,CMV16,CFLV17,GMT17,LGMS18}.
However, in these systems the underlying random variables have
long-range correlations, whereas in our problem the underlying random
variables are completely uncorrelated!  Thus the mechanism of the
first-order phase transition in our model is quite different from that
of the Coulomb gas systems studied
in~\cite{VMB08,MNSV09,VMB10,NMV10,NMV11,MNSV11,TM13,MS14,CMV16,CFLV17,GMT17,LGMS18}. Here
we find a condensation transition analogous to that of mass transport
models~\cite{MEZ05,EMZ06,EH05,S08-leshouches,SEM14,SEM14b,SEM17}.\\

Finally, we have presented here only results for the case of external
field $E>0$, although we also have preliminar results for the case
$E=0$. We already know that the limit $E\rightarrow 0$ is singular:
the exponents controlling the asymptotic decay of $P(X,N)$ in the case
of zero external field are different from the finite field case. All
the details on $P(X,N)$'s large deviation form in the case of $E=0$
are going to be presented elsewhere~\cite{GM-E0}. The results of the
present paper also have important implications for an equilibrium
thermodynamics study of wave-function localization in the nonlinear
Schr\"odinger equation: this is the subject of another forthcoming
work~\cite{GM-DNLS}.

\section*{Acknowledgments}

We thank E. Bertin, F. Corberi, A. Puglisi and G. Schehr for useful
discussions. \mycol{We also warmly thank N. Smith for pointing out an
  algebraic error in~\ref{app2} in the previous version of the
  manuscript and for suggesting an argument for computing $r_c$, which
  is now reported in~\ref{app2:zc}}. G.G. acknowledges Financial
support from the Simons Foundation grant No.~454949 (Giorgio
Parisi). G.G. aknowledges LIPhy, Universit\`e Grenoble-Alpes, for kind
hospitality during the first stages of this work (support from ERC
Grant No.~ADG20110209, Jean-Louis Barrat).

\newpage

\appendix

\section{Asymptotic tails of $\mP(x)$}
\label{Asymp_Px}

In this Appendix, we present the asymptotic bahaviors of
the distribution of the displacement in a single run, namely the
marginal distribution $\mP(x)$ written in Eq.~(\ref{x_marg.1}) of
Sec.\ref{sec:model-results}. We first consider $p(\tau)=e^{-\tau}\,\theta(\tau)$ and $q(v)=e^{-v^2/2}/\sqrt{2\pi}$.
Let us first define the mean and the
variance of $\mP(x)$, which can be easily computed.  The mean is given
by \be \langle x\rangle= \langle v\rangle \langle \tau\rangle +
\frac{1}{2}\, E\, \langle \tau^2\rangle= E\, .
\label{mean_x}
\ee 
Similarly, the second moment is simply, 
\be 
\langle x^2\rangle =  \langle v^2\rangle \langle \tau^2\rangle + E 
\langle v\rangle \langle \tau^3\rangle + \frac{E^2}{4} \langle 
\tau^4\rangle  = 2+ 6\,E^2 
\ee 
and hence the variance is given by 
\be 
\sigma^2  = \langle x^2\rangle- \langle x\rangle^2= 2+ 5\, E^2
\label{var_x} \, . 
\ee 
To compute the full marginal distribution $\mP(x)$, we 
perform the Gaussian integral 
over $v$ to get 
\be 
\mP(x)= \frac{1}{\sqrt{2\pi}} \int_0^{\infty}
\frac{d\tau}{\tau}\,
\exp\left[-\tau- \frac{(x- E \tau^2/2)^2}{2\tau^2}\right]\, . 
\label{Px_marg}
\ee 
This integral is hard to compute exactly. However, we are only 
interested in the large $|x|$ asymptotic tails of $\mP(x)$.

To derive the asymptotics of $\mP(x)$ in Eq. (\ref{Px_marg}), it is
first convenient to rewrite it as
\be
\mP(x)= \frac{1}{\sqrt{2\pi}}\, e^{x\,E/2}\, \int_0^{\infty} 
\frac{d\tau}{\tau} \exp\left[-\tau-\frac{x^2}{2\tau^2}- 
\frac{E^2\tau^2}{8}\right]\, .  
\label{EP.0}
\ee
Since $\mP(x)$ is manifestly asymmetric, let us consider the two
limits $x\to \infty$ and $x\to -\infty$ separately.
Consider first the positive side $x\ge 0$. Let us first rescale 
$\tau= \sqrt{x}\, y$ in Eq. (\ref{EP.0}), and rewrite the integral
for any $x\ge 0$ as
\be
\mP(x)= \frac{1}{\sqrt{2\pi}} \int_0^{\infty}
\frac{dy}{y}\, \exp\left[- \sqrt{x}\, y- 
\frac{x}{2}\left(\frac{Ey}{2}-\frac{1}{y}\right)^2 \right]\, .
\label{EP.1}
\ee
This is a convenient starting point for analysing the asymptotic tail  
$x\to \infty$.
The dominant contribution to this integral for large $x$ comes from the
vicinity of $y=y^*=\sqrt{2/E}$ that minimizes the square inside the 
exponential. Setting $y= \sqrt{2/E} + z$, expanding around $z=0$ 
(keeping terms up to $O(z^2)$) and performing the resulting
Gaussian integration gives, to leading order for large positive $x$
\be
\mP(x) \approx \frac{1}{E} e^{2/E^2}\, x^{-1/2}\, e^{-\sqrt{2x/E}}\ .
\label{EP.2}
\ee

Turning now to the large negative $x$, we set $x=-|x|$ in Eq. 
(\ref{EP.0}) and rewrite it, for $x<0$ as
\begin{eqnarray}
\mP(x<0) &= e^{-E |x|/2} \int_0^{\infty} \frac{d\tau}{\tau}\,  
\exp\left[-\tau-\frac{|x|^2}{2\tau^2}-
\frac{E^2\tau^2}{8}\right] \nonumber \\
&= e^{-E\, |x|} \mP(|x|)
\label{EP.3}
\end{eqnarray}
where, in the second line, we used the expression of $\mP(x)$ in Eq. 
(\ref{EP.0}) with argument $|x|>0$. Hence, for large $x\to -\infty$, we can use the 
already 
derived asymptotics of $\mP(x)$ for large positive $x$ in Eq. 
(\ref{EP.2}). This then gives, to leading order as $x\to -\infty$,
\be
\mP(x)\approx \frac{1}{E} e^{2/E^2}\, e^{-E\,|x|} |x|^{-1/2}\, 
e^{-\sqrt{2|x|/E}} \, .
\label{EP.4}
\ee
The results in Eqs. (\ref{EP.2}) and (\ref{EP.4}) can then be combined
into the single expression
\begin{eqnarray}   
\mP(x) \approx \begin{cases}
c_1\, 
|x|^{-1/2}~e^{-(2/E)^{1/2}\, 
|x|^{1/2} }, \quad\quad\quad\,\, x\to \infty \\
c_1\,
|x|^{-1/2}~e^{-E\, |x|\, -(2/E)^{1/2}\,
|x|^{1/2} }, \quad x\to -\infty 
\end{cases}
\label{eq:mG-asymptotic}
\end{eqnarray}
where $c_1= e^{2/E^2}/E$ is a constant.  Thus the marginal PDF of $x$
has stretched exponential tails on both sides with stretching exponent
$\alpha=1/2$, but in addition on the negative side it has an overall
multiplicative exponential factor $e^{-E\, |x|}$.  We note that this
model with a field $E>0$ has been studied earlier
in~\cite{GP86,AP10,MALB12,GMPS12,GSPT13} under the name of Stochastic
Lorentz gas, but no investigation was carried out on its condensation
transitions and the associated first-order phase transitions.

{\red{Consider now the case where $E>0$, $p(\tau)=e^{-\tau}\,\theta(\tau)$, but
the velocity distribution $q(v)$ is bimodal as in Eq. (\ref{bimodal.1}).
Substituting $q(v)$ in Eq. (\ref{x_marg.1}) and carrying out the $v$ integration
gives
\be
\mP(x)= \frac{1}{2}\int_0^{\infty} d\tau\, e^{-\tau}\, \left[\delta\left(x-v_0\tau-\frac{E}{2}\tau^2\right)+
\delta\left(x+v_0\tau-\frac{E}{2}\tau^2\right)\right]\, .
\label{bimod1.A1}
\ee
Now, for large $x>0$, the leading contribution comes from large $\tau$, hence one can neglect $v_0\tau$ terms leading to
\begin{eqnarray}
\mP(x) &\approx & \int_0^{\infty} d\tau\, e^{-\tau}\, \delta\left(x- \frac{E}{2}\tau^2\right) \nonumber \\
&=& \frac{1}{\sqrt{2\,E\,x}}\, e^{-\sqrt{2x/E}}\, .
\label{bimod2.A1}
\end{eqnarray}
Hence, for large $x>0$, the marginal distribution $\mP(x)$ has a stretched exponential decay and it satisfies the
criterion for positive condensation. In contrast, for negative $x$ and $E>0$, it is easy to see that $\mP(x)$ strictly
vanishes for $x< -v_0^2/{2E}$. Thus the marginal distribution is bounded on the negative side. Consequently it does not
satisfy the condensation criterion for negtaive $x$. Thus, in this example, we only have one sided condensation
in the displacement PDF $P(X,N)$. 

We next consider the standard RTP case: $E=0$, $p(\tau)=e^{-\tau}\theta(\tau)$ and $q(v)$ bimodal as in Eq. (\ref{bimodal.1}).
Substituting $q(v)$ in Eq. (\ref{x_marg.1}) and carrying out the $v$ integral gives
\begin{eqnarray}
\mP(x)&=& \frac{1}{2} \int_0^{\infty} d\tau\, e^{-\tau}\, \left[\delta(x-v_0\tau)+ \delta(x+v_0\tau)\right] \nonumber \\
&=& \frac{1}{2v_0}\, e^{-|x|/v_0}\, .
\label{rtp_standard.A1}
\end{eqnarray}
Thus, in this case, the marginal $\mP(x)$ decays exponentially on both sides and hence does not satisfy
the condensation criterion. Consequently, for the standard RTP, we do not have condensation on either side.
However, for $E=0$, and arbitrary velocity distribution $q(v)$ with a finite width,
the condensation transition is restored~\cite{GM-E0}.

}}

\section{Derivation of the rate function $\chi(z)$ in the intermediate
matching regime}
\label{app2}

In this Appendix we study the leading large $N$ behavior of the
integral that appears in the expression for $P_{\textrm{A}}(z,N)$ in
Eq. (\ref{eq:PX-matching}):
\be
I_N(z)= \int_{\Gamma_{(+)}} ds\frac{1}{\sqrt{s}}~e^{N^{1/3}F_z(s)}
%~\approx~e^{-N^{1/3}\chi(z)},
\label{eq:anomalous-int-app}
\ee
where $z\ge 0$ can be thought of as a parameter and 
\be
F_z(s) = sz + \frac{1}{2}\,\sigma^2\,s^2 +\frac{1}{2sE} \, ,
\label{Fzs.1}
\ee
with $\sigma^2=2+5 E^2$. It is important to recall that the contour
$\Gamma_{(+)}$ is along a vertical axis in the complex $s$-plane with
its real part negative, i.e. ${\rm Re}(s)<0$.  Thus, we can deform
this contour only in the upper left quadrant in the complex $s$ plane
(${\rm Re}(s)<0$ and ${\rm Im}(s)>0$), but we can not cross the branch
cut on the real negative axis, nor can we cross to the $s$-plane where
${\rm Re}(s)>0$. \blue{A convenient choice of the deformed contour, as we will
see shortly, is the $\Gamma_{(+)}$ rotated anticlockwise by an angle $\pi/2$, so that
the contour now goes along the real negative $s$ from $0$ to $-\infty$.}

\begin{figure}
  \includegraphics[width=\columnwidth]{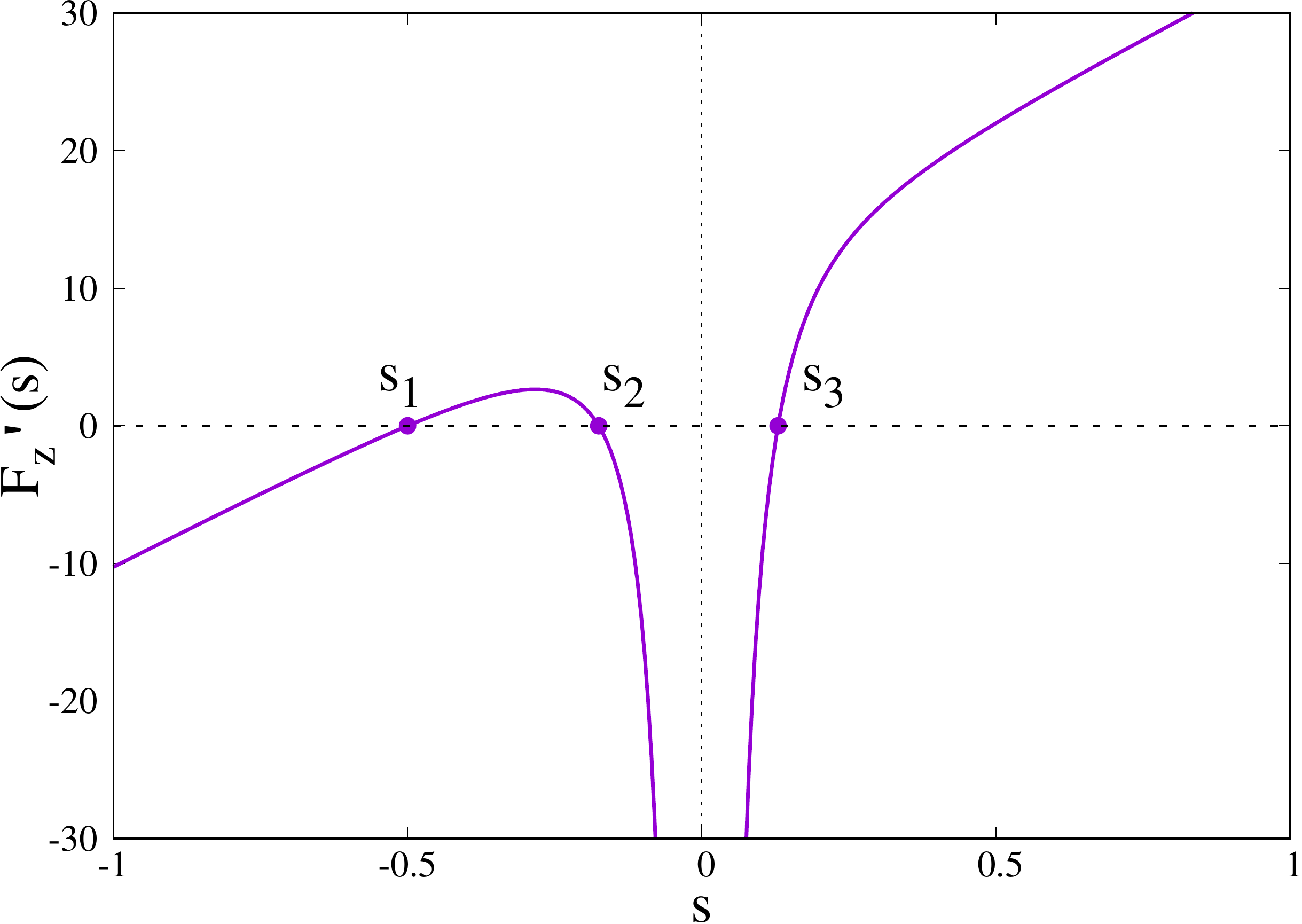}
\caption{A plot of $F_z'(s)=z+ \sigma^2\, s- \frac{1}{2Es^2}$ as a function of $s$ ($s$ 
real) for $z=12$, $E=2$ and $\sigma^2=2+5E^2=22$. There are three zeros on the real $s$
axis (obtained by \emph{Mathematica}) at $s_1=-0.5$, $s_2=-0.175186\dots$ and 
$s_3=0.129732\dots$.}
\label{fig:der}
\end{figure}

To evaluate the integral in Eq. (\ref{eq:anomalous-int-app}), it is natural
to look for a saddle point of the integrand in the complex $s$ plane in
the left upper quadrant, with fixed $z$. 
Hence, we look for solutions for the stationary points of the function $F_z(s)$
in Eq. (\ref{Fzs.1}). They are given by the zeros of the cubic equation
\begin{equation}
F_z'(s)= \frac{dF_z(s)}{ds}= z+ \sigma^2\, s- \frac{1}{2Es^2} \equiv 0
\label{dfzs.1}
\end{equation}
%%
%5
As $z\ge 0$ varies, the three roots move in the complex $s$ plane.  It
turns out that for $z<z_l$ (where $z_l$ is to be determined), there is
one positive real root and two complex conjugate roots.  For example,
when $z=0$, the three roots of Eq. (\ref{dfzs.1}) are respectively at
$s=(2E\sigma^2)^{-1/2} \, e^{i\phi}$ with $\phi=0$, $\phi=2\pi/3$ and
$\phi=4\pi/3$.  However, for $z>z_l$, all the three roots collapse on
the real $s$ axis, with $s_1<s_2<s_3$. The roots $s_1<0$ and $s_2<0$
are negative, while $s_3>0$ is positive.  For example, in
Fig. (\ref{fig:der}), we plot the function $F_z'(s)$ in
Eq. (\ref{dfzs.1}) as a function of real $s$, for $z=12$ and $E=2$ (so
$\sigma^2=2+5 E^2=22$).  One finds, using Mathematica, three roots at
$s_1= -1/2$ (the lowest root on the negative side),
$s_2=-0.175186\dots$ and $s_3= 0.129732\dots$.
We can now determine $z_l$ very easily. As $z$ decreases, the two
negative roots $s_1$ and $s_2$ approach each other and become
coincident at $z=z_l$ and for $z<z_l$, they split apart in the complex
$s$ plane and become complex conjugate of each other, with their real
parts identical and negative. When $s_1<s_2$, the function $F_z'(s)$
has a maximum at $s_m$ with $s_1<s_m<s_2$ (see Fig.~\ref{fig:der}). As $z$
approaches $z_l$, $s_1$ and $s_2$ approach each other, and
consequently the maximum of $F_z'(s)$ between $s_1$ and $s_2$ approach
the height $0$. Now, the height of the maximum of $F_z'(s)$ between
$s_1$ and $s_2$ can be easily evaluated.  The maximum occurs at
$s=s_m$ where $F_z''(s)=0$, i.e, at $s_m=- (E\sigma^2)^{-1/3}$.  Hence
the height of the maximum is given by
\begin{equation}
F_z'(s=s_m)= z+ \sigma^2\, s_m + \frac{1}{2s_m E}= z- 
\frac{3}{2}\left(\frac{\sigma^4}{E}\right)^{1/3}\, .
\label{F_zmax.1}
\end{equation}
Hence, the height of the maximum becomes exactly zero when 
\begin{equation}
z=z_l= \frac{3}{2}\left(\frac{\sigma^4}{E}\right)^{1/3}\, .
\label{zl_exact.1}
\end{equation} 
Thus we conclude that for $z>z_l$, with $z_l$ given exactly in
Eq.~(\ref{zl_exact.1}), the function $F_z'(s)$ has three real roots at
$s=s_1<0$, $s_2 <0$ and $s_3>0$, with $s_1$ being the smallest
negative root on the real axis. For $z<z_l$, the pair of roots are
complex (conjugates). However, it turns out (as will be shown below)
that for our purpose, it is sufficient to consider evaluating the
integral in Eq.  (\ref{eq:anomalous-int-app}) only in the range
$z>z_l$ where the roots are real and evaluating the saddle point
equations are considerbaly simpler.  So, focusing on $z>z_l$, out of
these $3$ roots as possible saddle points of the integrand in Eq.
(\ref{eq:anomalous-int-app}), we have to discard $s_3>0$ since our
saddle points have to belong to the upper left quadrant of the complex
$s$ plane. This leaves us with $s_1<0$ and $s_2<0$.  \blue{Now, we
  deform our vertical contour $\Gamma_{(+)}$ by rotating it
  anticlockwise by $\pi/2$ so that it runs along the negative real
  axis. Between the two stationary points $s_1$ and $s_2$, it is easy
  to see (see Fig. (\ref{fig:der})) that $F_z''(s_1)>0$ (indicating
  that it is a minimum along real $s$ axis) and $F_z''(s_2)<0$
  (indicating a local maximum).  Since the integral along the deformed
  contour is dominated by the maximum along real negative $s$ for
  large $N$, we should choose $s_2$ to be the correct root, i.e., the
  largest among the negative roots of the cubic equation $z+\sigma^2
  s- 1/(2Es^2)=0$.}

Thus, evaluating the integral at $s^*=s_2$ (and discarding
pre-exponential terms) we get for large $N$
\begin{equation}
I_N(z) \approx \exp[- N^{1/3} \chi(z)]\
\label{saddle_s1.1}
\end{equation}
where the rate function $\chi(z)$ is given by
\begin{equation}
\chi(z) = - F_z(s=s_2)=-s_2\,z-\frac{1}{2}\sigma^2 s_2^2- \frac{1}{2s_2 E}
\label{chiz.B1}
\end{equation}
The right hand side can be further simplified by using
the saddle point equation (\ref{dfzs.1}), i.e.,
$z+ \sigma^2 s_2- 1/{2Es_2^2}=0$. This gives
\begin{equation}
\chi(z)= - \frac{zs_2}{2}-\frac{3}{4Es_2}\, .
\label{chiz.2}
\end{equation}

\subsection{Asymptotic behavior of $\chi(z)$}
\label{app2:perturbative}

We now determine the asymptotic behavior of the rate function
$\chi(z)$ in the range $z_l<z<\infty$, where $z_l$ is given in
Eq. (\ref{zl_exact.1}).  Essentially, we need to determine $s_2$ (the
largest among the negative roots) as a function of $z$ by solving
Eq. (\ref{dfzs.1}), and substitute it in Eq. (\ref{chiz.2}) to
determine $\chi(z)$.

We first consider the limit $z\to z_l$ from above, where $z_l$ is
given in Eq. (\ref{zl_exact.1}). As $z\to z_l$ from above, we have
already mentioned that the two negative roots $s_1$ and $s_2$ approach
each other. Finally at $z=z_l$, we have $s_1=s_2=s_m$ where $s_m=-
(E\sigma^2)^{-1/3}$ is the location of the maximum between $s_1$ and
$s_2$.  Hence as $z\to z_l$ from above, $s_2\to s_m= -
(E\sigma^2)^{-1/3}$.  Substituting this value of $s_2$ in
Eq. (\ref{chiz.2}) gives the limiting behavior
\begin{equation}
\chi(z) \to \frac{3}{2}\, \left(\frac{\sigma}{E}\right)^{2/3}\quad {\rm as}\,\, z\to z_l
\label{chiz_ll}
\end{equation}
as announced in the first line of Eq. (\ref{eq:chi-asymptotics-intro}).

To derive the large $z\to \infty$ behavior of $\chi(z)$ as announced in the second line
of Eq. (\ref{eq:chi-asymptotics-intro}), it is first convenient
to re-parametrize $s_2$ and define
\begin{equation}
s_2= -\frac{1}{\sqrt{2Ez}} \theta_z \,.
\label{eq:rescale-sz}
\ee
Substituting this in Eq. (\ref{dfzs.1}), it is easy to see that $\theta_z$ satisfies
the cubic equation
\be
-b(z)~\theta_z^3+\theta_z^2-1 =0,
\label{eq:sad-eq-pert}
\ee
where
\be
b(z) = \frac{\sigma^2}{\sqrt{2E}} \frac{1}{z^{3/2}}.
\label{eq:bz-def}
\ee
Note that due to the change of sign in going from $s_2$ to $\theta_z$,
we now need to determine the smallest {\em positive} root of
$\theta_z$ in Eq. (\ref{eq:sad-eq-pert}).  In terms of $\theta_z$,
$\chi(z)$ in Eq. (\ref{chiz.2}) reads
\begin{equation}
\chi(z)= ~\frac{\sqrt{z}}{2\sqrt{2 E}}~\frac{\theta_z^2+3}{\theta_z}\, .
\label{eq:chi-useful}
\end{equation}

The formulae in Eqs.~(\ref{eq:sad-eq-pert}), (\ref{eq:bz-def})
and (\ref{eq:chi-useful}) are now particularly suited for the large
$z$ analysis of $\chi(z)$. From
Eqs.~(\ref{eq:sad-eq-pert}),(\ref{eq:bz-def}) it follows that in the
limit $z\to \infty$ we have that $b(z)\to 0$, so that $\theta_z\to 1$. Hence,
for large $z$ or equivalently small $b(z)$, we can obtain a
perturbative solution of Eq. (\ref{eq:sad-eq-pert}). To leading order,
it is easy to see that
\begin{equation}
\theta_z= 1+ \frac{b(z)}{2} + \mathcal{O}\left({b(z)}^2\right)\, .
\label{pert_thetaz.1}
\end{equation}
with $b(z)$ given in Eq.~(\ref{eq:bz-def}).  Substituting this in
Eq.~(\ref{eq:chi-useful}) gives the large $z$ behavior of $\chi(z)$
\be
\chi(z) = \sqrt{\frac{2}{E}}~\sqrt{z}-\frac{\sigma^2}{4E}\frac{1}{z} + \mO\left(\frac{1}{z^{5/2}}\right).
\label{eq:rate-function-with-theta}
\ee
as announced in the second line of Eq. (\ref{eq:chi-asymptotics-intro}).

\mycol{
\subsection{Explicit expression of $\chi(z)$}
\label{app2:exact}

While the excercises in the previous subsections were instructive, it
is also possible to obtain an explicit expression for $\chi(z)$ by
solving the cubic equation (\ref{eq:sad-eq-pert}) with {\it
  Mathematica}. The smallest positive root of
Eq.~(\ref{eq:sad-eq-pert}), using {\it Mathematica}, reads
%%
%\begin{widetext} 
\begin{equation} 
\theta_z =  \frac{1}{3b_z} + \frac{1}{3\cdot 2^{2/3} b_z} \frac{(1-i\sqrt{3})}{\left( -2
      + 27 b_z^2 + 3 \sqrt{-12+81 b_z^2} \right)^{1/3}}  + 
\frac{1}{3\cdot 2^{4/3} b_z}(1+i\sqrt{3}) \left( -2 + 27 b_z^2 + 3 \sqrt{-12+81 b_z^2} \right)^{1/3}
\label{thetaz_math1}
\end{equation}
%%
%\end{widetext}
where $b_z$, used as an abbreviation for $b(z)$, is given in
Eq.~(\ref{eq:bz-def}). Using the expression of $z_l$ in
Eq.~(\ref{zl_exact.1}), we can re-express $b_z$ conveniently in a
dimensionless form
\begin{equation} 
b_z^2 = \frac{1}{2}~\left( \frac{2}{3} \frac{z_l}{z} \right)^3 \, .
\end{equation}
Consequently, the solution $\theta_z$ in Eq. (\ref{thetaz_math1}) in
terms of the adimensional parameter $r=z/z_l\ge 1$ reads as
\begin{equation}  
\theta_z \equiv \theta(r)= \frac{\sqrt{3}}{4} r^{3/2} \left[ 2 + \frac{(1-i\sqrt{3})}{g(r)} +
  (1+i\sqrt{3})g(r) \right]  
\end{equation}
where 
\begin{equation} 
g(r) = \frac{1}{r} \left( 1 + i ~\sqrt{r^3-1} \right)^{2/3}. 
\end{equation}
By multiplying both numerator and denominator of $\theta(r)$ by $(1-i
~\sqrt{r^3-1})^{2/3}$
one ends up, after a little algebra, with the following expression
\begin{equation} 
\theta(r) = \frac{\sqrt{3}}{4} r^{3/2} \left[ 2 + \frac{1}{r}\left(
  \xi~\zeta_r^{2/3}+\overline{\xi}~\overline{\zeta}_r^{2/3} \right)  \right],
\label{eq:thetar-complex}
\end{equation}
where $\xi$ and $\zeta_r$ denotes, respectively, a complex number and a
complex function of the real variable $r$:
\begin{eqnarray}
\xi &=& 1 + i \sqrt{3} \nonumber \\
\zeta_r &=& 1 + i ~\sqrt{r^3-1},
\label{eq:complex-exp}
\end{eqnarray}
and we have also introduced the related complex conjugated quantities:
\begin{eqnarray}
\overline{\xi} &=& 1 - i \sqrt{3} \nonumber \\
\overline{\zeta}_r &=& 1 - i ~\sqrt{r^3-1},
\label{eq:complex-exp-2}
\end{eqnarray}

We can then write the complex expressions in
Eq.~(\ref{eq:thetar-complex}) both in their polar form, i.e., $\zeta_r =
\rho_re^{i\phi_r}$ and $\xi = \rho e^{i\phi}$, with, respectively:
\begin{eqnarray}
  \rho_r &=& r^{3/2} \nonumber \\
  \phi_r &=& \textrm{arctan}(\sqrt{r^3-1})
\label{eq:args-zetar}
\end{eqnarray}
and
\begin{eqnarray}
  \rho &=& 2 \nonumber \\
  \phi &=& \textrm{arctan}(\sqrt{3}) = \frac{\pi}{3}.
  \label{eq:args-xi}
\end{eqnarray}
Finally, by writing $\xi$ and $\zeta_r$ inside
Eq.~(\ref{eq:thetar-complex}) in their polar form and taking advantage
of the expressions in Eqns.~(\ref{eq:args-zetar}),(\ref{eq:args-xi}) we get:
\begin{eqnarray} 
  \theta(r) &=& \frac{\sqrt{3}}{4} r^{3/2} \left[ 2 + \frac{1}{r}~\rho~\rho_r^{2/3} \left( e^{i\left( \phi + \frac{2}{3} \phi_r \right)} + e^{-i\left( \phi + \frac{2}{3} \phi_r \right)} \right)  \right] = \nonumber \\ 
  &=& \frac{\sqrt{3}}{2} r^{3/2} \left[ 1 + 2 \cos\left( \frac{\pi}{3} + \frac{2}{3} \textrm{arctan}(\sqrt{r^3-1})\right) \right] 
\label{eq:theta-r}
\end{eqnarray}
In order to draw explicitly the function $\chi(z)$, e.g. with the help
of \emph{Mathematica}, one can plug the expression of $\theta(r=z/z_l)$ from
Eq.~(\ref{eq:theta-r}) into the following formula:
\begin{equation}
\chi(z)=\frac{\sqrt{z}}{2\sqrt{2E}}~\frac{\theta(z/z_l)^2+3}{\theta(z/z_l)}\, ,
\label{chiz.B3}
\end{equation}
}

\mycol{
\subsection{The critical value $z_c$}
\label{app2:zc}

We show here how to compute the critical value $z_c$ at which
$\chi(z)$ equals $z^2/(2\sigma^2)$, i.e., the value at which the two
branches in Fig.~\ref{fig:transition} cross each other.  To make the
computations easier, it is convenient to work with dimensionless
variables.  Using $z_l= (3/2) (\sigma^4/E)^{1/3}$ from
Eq.~(\ref{zl_exact.1}), we express $z$ in units of $z_l$, i.e., we
define
\begin{equation}
r= \frac{z}{z_l}= \frac{2z}{3}\, \left(\frac{E}{\sigma^4}\right)^{1/3}\, .
\label{r_def}
\end{equation}
In terms of $r$, one can rewrite $b(z)$ in Eq.~(\ref{eq:bz-def}) as
(using the shorthand notation $b_z=b(z)$):
\begin{equation}
b_z^2= \frac{1}{2}\, \left(\frac{2}{3r}\right)^{3}\, .
\label{b_z}
\end{equation}
Consequently, Eq.~(\ref{eq:sad-eq-pert}) reduces to
\begin{equation}
-\frac{1}{\sqrt{2}}\, \left(\frac{2}{3}\right)^{3/2}\, r^{-3/2}\, \theta(r)^3 + \theta(r)^2 -1=0\, ,
\label{theta_r}
\end{equation}
%%%
where $\theta(r)= \theta_{z= r z_l}$ is dimensionless. Quite
remarkably, it turns out that to determine the critical value $z_c$,
rather conveniently we do not need to solve the above cubic equation,
Eq.~(\ref{theta_r}).  Indeed, at $z=z_c$, i.e., $r=r_c$, equating
$\chi(z_c)= z_c^2/{2\sigma^2}$, we get
%%%
\begin{equation}
\frac{\sqrt{z_c}}{2\sqrt{2E}}\,\left[ \frac{\theta(r_c)^2 +3}{\theta(r_c)}\right]= \frac{z_c^2}{2\sigma^2}\, .
\label{zc.1}
\end{equation}
Expressing in terms of $r_c$, Eq. (\ref{zc.1}) simplifies to
\begin{equation}
\frac{\theta^2(r_c)+3}{\theta(r_c)}= \frac{3^{3/2}}{2}\, r_c^{3/2}\, .
\label{rc.1}
\end{equation}
Consider now Eq.~(\ref{theta_r}) evaluated at $r=r_c$. In this equation, we replace $r_c$ by its
expression in Eq.~(\ref{rc.1}). This immediately gives $\theta(r_c)^2=3/2$ and hence
\begin{equation}
\theta(r_c)= \sqrt{\frac{3}{2}}\, .
\label{thetarc}
\end{equation}
Using this exact $\theta(r_c)$ in Eq. (\ref{rc.1}) gives
\begin{equation}
r_c= \frac{z_c}{z_l}= 2^{1/3}= 1.25992\dots
\label{exact_rc}
\end{equation}
It is now straightforward to check that the expression of $\theta(r)$
written in Eq.~(\ref{eq:theta-r}) is consistent with the result just
found, i.e., from it we retrieve $\theta(r_c=2^{1/3})=\sqrt{3/2}$. We
have that
\begin{eqnarray}
  \theta(r_c=2^{1/3}) &=& \frac{\sqrt{3}}{2} r_c^{3/2} \left[ 1 + 2 \cos\left( \frac{\pi}{3} + \frac{2}{3} \textrm{arctan}(\sqrt{r_c^3-1})\right) \right] = \nonumber \\
  &=& \sqrt{\frac{3}{2}} \left[ 1 + 2 \cos\left( \frac{\pi}{3} + \frac{2}{3} \textrm{arctan}(1)\right) \right] = \sqrt{\frac{3}{2}} \left[ 1 + 2 \cos\left( \frac{\pi}{2}\right) \right] = \nonumber \\
  &=& \sqrt{\frac{3}{2}},
  \end{eqnarray}
as expected. \\

For comparison to numerical simulations, we chose $E=2$, for which
$\sigma^2= 2+5E^2= 22$.  We get $z_l= (3/2) (\sigma^4/E)^{1/3}=
9.34752\ldots$, which gives $z_c= r_c z_l= (1.25992\dots) z_l=
11.7771\dots.$ This is represented by a black dotted vertical line in
Fig.~\ref{fig:numerics}.}


\begin{thebibliography}{38}
\bibitem{complex-saddle} The general formula for the saddle-point
  approximation of a contour integral in the complex plane which
  depend on a large parameter $\lambda$ is \be \mI(\lambda) =
  \int_\Gamma dz~f(z)~e^{\lambda S(z)} = \left(\frac{2\pi}{-\lambda
    S''(z_0)}\right)^{\frac{1}{2}}~e^{\lambda S(z_0)}~f(z_0) \ee
\end{thebibliography}
\end{document}